\documentclass[12pt]{article}

\usepackage{a4wide, amsmath, latexsym, epsfig,
  psfrag,graphicx, rotating, fancyhdr, tabularx, afterpage}

\begin{document}

\thispagestyle{empty}
\setcounter{page}{0}
\def\thefootnote{\fnsymbol{footnote}}

\begin{flushright}
MPP-2006-117 \\
Freiburg-THEP 06/16\\
hep-ph/0610267 \\
\end{flushright}

\vspace{1cm}

\begin{center}

{\large\sc {\bf One-loop calculations \\
\vspace{0.4cm}
of the decay of the next-to-lightest neutralino in the MSSM}} 

\vspace{1cm}

{\sc Manuel Drees$^{1}$
\footnote{email: drees@th.physik.uni-bonn.de},
~Wolfgang Hollik$^{2}$
\footnote{email: hollik@mppmu.mpg.de}
~and {\sc Qingjun Xu}$^{2}$}
\footnote{email: qingjun.xu@physik.uni-freiburg.de\\
\hspace*{3.5ex}Present address: Physikalisches Institut, 
Albert-Ludwigs-Universit\"at Freiburg\\
\hspace*{21.ex}Hermann-Herder-Str. 3, D-79104 Freiburg, Germany}

\vspace*{1cm}
{\sl
$^1$Physikalisches Institut der Universit\"at Bonn\\
 Nussallee 12, D-53115 Bonn, Germany 

\vspace*{0.4cm}

$^2$Max-Planck-Institut f\"ur Physik (Werner-Heisenberg-Institut)\\
F\"ohringer Ring 6, D-80805 Munich, Germany}
      
\end{center}

\vspace*{1cm}

\begin{abstract}
\noindent
We calculate one-loop corrections to the decays of the next-to-lightest
neutralino $\tilde{\chi}_2^0$ into the lightest neutralino $\tilde{\chi}_1^0$
and two leptons; this includes diagrams where a real photon is emitted. In
cases where two-body decays $\tilde{\chi}_2^0 \rightarrow \tilde{l}^\pm_1
l^\mp \rightarrow \tilde{\chi}_1^0 l^- l^+$ are kinematically allowed, we
calculate these decays both with and without the single-pole approximation,
and find consistent results. For example, for the minimal supergravity
parameter set SPS1a, the integrated partial widths (the branching ratios) 
for $\tilde{\chi}_2^0\rightarrow \tilde{\chi}_1^0 l^- l^+\ (l = e, \mu)$ 
are enhanced by about 15.5 (13.6) percent by the one-loop corrections. 
We also study a scenario where $\tilde \chi_2^0$ cannot undergo two-body 
decays, and find corrections to these branching ratios of about 13.6 percent. 
Moreover, we study the dilepton invariant mass ($M_{l^+ l^-}$) distribution, 
whose endpoint is often used in analyses that aim to reconstruct 
(differences of) supersymmetric particle masses at the LHC. The shape of this 
distribution is altered significantly by the emission of hard photons. For 
example, for the SPS1a parameter set the peak of the $M_{l^+ l^-}$ distribution
is shifted by several GeV when these contributions are included. 

\end{abstract}

\def\thefootnote{\arabic{footnote}}
\setcounter{footnote}{0}

\newpage

\section{Introduction}\label{introduction}

Supersymmetry (SUSY) \cite{MSSM} is one of the best motivated extensions of
the Standard Model (SM) of particle physics. If SUSY exists at the electroweak
scale, experiments at future high energy colliders should be able to discover
the superpartners of known particles, and to study their properties \cite{LHC,
  TESLA}. From the precise measurement of the masses, production cross
sections and decay branching ratios of these superpartners, the fundamental
parameters of the underlying SUSY models can be determined. This will help us
to reconstruct the SUSY breaking mechanism.

In the Minimal Supersymmetric Standard Model (MSSM) \cite{MSSM} with conserved
$R$-parity, the lightest supersymmetric particle (LSP), which in many scenarios
is the lightest neutralino $\tilde{\chi}_1^0$, appears at the end of the decay
chain of each supersymmetric particle. The LSP escapes the detector, giving
the characteristic SUSY signature of missing energy. While this helps to
suppress backgrounds from SM processes, it also makes the measurement of
supersymmetric particle masses more difficult. This is true in particular 
at hadron colliders like the LHC, where the total energy in a given partonic 
collision is not known. 

At the LHC, the total SUSY production cross section is expected to be
dominated by the production of gluinos and squarks, which decay into lighter
charginos or neutralinos. Of particular interest are decay chains leading to
the next-to-lightest neutralino $\tilde{\chi}_2^0$.  $\tilde \chi_2^0$ in
turn can always undergo the three-body decays $\tilde \chi_2^0 \rightarrow
\tilde\chi_1^0 f \bar f$, at least for light SM fermions $f$. Depending on
the neutralino, sfermion and Higgs boson masses, the two-body decays $\tilde
\chi_2^0 \rightarrow \tilde f \bar f \rightarrow \tilde \chi_1^0 f \bar f$
and/or $\tilde \chi_2^0 \rightarrow \tilde \chi_1^0 Z/\phi \rightarrow
\tilde \chi_1^0 f \bar f$ may also be open, where $\phi$ stands for one of the
three neutral Higgs bosons of the MSSM; of course, Higgs intermediate states
will contribute negligibly if $f = e$ or $\mu$. These leptonic final states
are of particular interest, since they can be identified relatively easily
even at the LHC. Moreover, the dilepton invariant mass distribution can be
measured accurately. In particular, the endpoint of this distribution is used
in several analyses that aim to reconstruct (differences of) supersymmetric 
particle masses \cite{LHC, LHC/LC}. Under favorable circumstances it has been 
shown that this endpoint can be measured to an accuracy of $0.1\%$ at the LHC
\cite{LHC}. In order to match this accuracy, at least one-loop corrections to
$\tilde \chi_2^0$ decays have to be included.

Turning to the planned $e^+e^-$ linear collider ILC, $\tilde \chi_1^0 \tilde
\chi_2^0$ production is often the first process that is kinematically
accessible \cite{ddk1} (other than $\tilde \chi_1^0$ pair production, which
leads to an invisible final state). The detailed analysis of $\tilde \chi_2^0$
decays can then yield information about heavier supersymmetric particles. 
Under favorable circumstances, ${\cal O}(10^4)$ $\chi_2^0 \rightarrow \chi_1^0 
l^+ l^-$ decays may be observed at the ILC, again making the inclusion of 
quantum corrections mandatory to match the experimental precision. In this 
paper we present a complete calculation of these corrections in the MSSM.

The general MSSM has more than one hundred unknown free parameters. Therefore,
it is not practicable to scan over the entire parameter space. Instead,
several ``benchmark scenarios'' have been suggested \cite{SPS1a}, which are
meant to illustrate characteristic features of various scenarios of
SUSY breaking. Among those, the so-called SPS1a parameter set, which
has been defined in the framework of the mSUGRA scenario \cite{MSSM}, has been
studied particularly widely. It gives rise to a particle spectrum where many
states are accessible both at the LHC and at a 500 GeV ILC
\cite{LHC/LC}. The masses of the relevant neutralinos and sleptons at this
benchmark point are listed in Table~\ref{spstab}. Note in particular that the
two-body decays $\tilde{\chi}_2^0\rightarrow \tilde{l}^\pm_1 l^\mp
\rightarrow \tilde{\chi}_1^0 l^- l^+$ are kinematically allowed; here $\tilde
l_1$ stands for the lighter one of the two charged sleptons of a given flavor. 
No other two-body decay mode is open. Moreover, squarks are so heavy that
non-leptonic $\tilde \chi_2^0$ decays can be neglected in this scenario. Note
that $\tilde \chi_1^0$ is mostly a $U(1)_Y$ gaugino (bino), while $\tilde
\chi_2^0$ is dominated by its neutral $SU(2)$ gaugino (wino) component; this
is typical for most scenarios where the gaugino mass unification relation
holds \cite{MSSM}. Leptonic two-body decays of $\tilde \chi_2^0$ have been
investigated at tree-level in Ref. \cite{LHC/LC}, three-body decays of
$\tilde{\chi}_2^0$ have been also studied at tree-level in
Refs. \cite{treedonea,noya}.

\begin{table}[htb] \label{spstab}
\begin{center}
\begin{tabular}{|c|c|c|c|c|c|c|c|c|}\hline
particle & $\tilde{\chi}_2^0$ & $\tilde{\chi}_1^0$ &
$\tilde{e}_1 \, (\tilde{\mu}_1)$ & $\tilde{e}_2 \, (\tilde{\mu}_2)$ &
$\tilde{\tau}_1$ & $\tilde{\tau}_2$ & $\tilde{\nu}_{e\, (\mu)}$ &
$\tilde{\nu}_{\tau}$ \\ \hline
mass [GeV] & 176.6 & 96.2 & 142.7 &  202.3 & 133.0 & 206.3 & 186.0 & 185.1  \\
\hline 
\end{tabular}
\end{center}
\caption{Masses of the relevant neutralinos and sleptons for parameter
  set SPS1a \cite{SPS1a}. }
\end{table}

In this paper, we calculate leptonic $\tilde \chi_2^0$ decays at one-loop
level. Cases where $\tilde \chi_2^0$ has two-body decays $\tilde\chi_2^0 
\rightarrow \tilde l_1^\pm l^\mp \rightarrow \tilde \chi_1^0 l^- l^+$
are treated both completely and in a single-pole approximation.
In the complete calculation one has to employ complex slepton masses in the 
relevant propagators and one-loop integrals. The single-pole approximation 
in this case is performed by treating $\tilde{\chi}_2^0$ decays as the 
production and decay of the sleptons $\tilde{l}_1$. We compare the results of
both methods, and find good agreement for the SPS1a parameter set. We also
analyze a scenario where $\tilde \chi_2^0$ only has three-body decays. In
addition to calculating the integrated partial widths, we study the
differential decay width of $\tilde{\chi}_2^0$ as a function of the dilepton
invariant mass. If $\tilde \chi_2^0$ can undergo two-body decay, the shape of
this distribution is essentially only affected by the emission of real
photons; as well known, these contributions have to be added to the one-loop
corrections to cancel infrared divergences. If $\tilde \chi_2^0$ only
undergoes three-body decays, the shape of this distribution is also altered
by the virtual corrections. In order to obtain the total decay width of 
$\tilde\chi_2^0$ and hence the branching ratios of its leptonic decays,
the invisible decays $\tilde\chi_2^0 \rightarrow \tilde \chi_1^0 \nu_l \bar 
\nu_l$ and the hadronic decays $\tilde\chi_2^0 \rightarrow \tilde \chi_1^0 
q \bar q$ are also calculated.

The paper is organized as follows. In Sec.~\ref{ReMSSM} we briefly
summarize the renormalization of those sectors of the MSSM which are relevant
for the decays of $\tilde{\chi}_2^0$. The calculation of the tree-level decay
widths is outlined in Sec.~\ref{tree}. Sec.~\ref{oneloop} discusses how
to calculate these decays at one-loop level, including the emission of real
photons. The complete one-loop calculation and the one-loop calculation in
the single-pole approximation are presented in Secs.
\ref{complete} and \ref{approximate}, respectively. In Sec.~\ref{brlepton} 
the total decay width of $\tilde\chi_2^0$ and the branching ratios of 
the leptonic decays are studied. Some numerical results
are given in Sec.~\ref{numerical}. We conclude our work in 
Sec.~\ref{conclusion}.

\section{Renormalization of the MSSM}\label{ReMSSM}

In order to calculate the higher-order corrections, one must renormalize the
parameters and the fields of the MSSM. Several approaches for the
renormalization of the MSSM have been developed~\cite{MSSMRE, others,
guasch, ReNeu, ReSle, ReHiggs, RetanB}. 
Here we employ on-shell renormalization
following the strategy of Refs.~\cite{ReNeu, ReSle},
This renormalization scheme is
convenient for our purposes, since it ensures that the relevant supersymmetric 
particle masses are (almost) the same at one-loop level as at tree level; in
particular, the endpoint of the $M_{ll}$ distribution is the same in both
cases. We assume here that all relevant parameters are real quantities; this
amounts to the assumption that the soft-SUSY-breaking terms conserve CP.

\subsection{Renormalization of the Chargino/Neutralino sector}

Loop corrections to the masses and mixing angles of charginos and neutralinos
were first discussed in Ref. \cite{earlyino}. The independent SUSY parameters
in the chargino/neutralino mass matrices are the electroweak gaugino mass
parameters $M_1$, $M_2$, and the supersymmetric Higgs mass parameter $\mu$.
These mass matrices also depend on the masses of the electroweak $W$ and $Z$
bosons, on the weak mixing angle $\theta_W$, and on the ratio of vacuum
expectation values (VEVs) $\tan\beta$; all these parameters are renormalized
independently from the chargino/neutralino sector, as outlined below. In order
to obtain finite $S$-matrix elements and Green's functions for chargino
fields, we introduce a counterterm for the chargino mass matrix $X$, as well
as field renormalization constants for the physical (mass eigenstate) four
component (Dirac) chargino fields $\tilde{\chi}^+_i (i =1, 2)$ \cite{ReNeu}:
\begin{eqnarray}
X &\longrightarrow & X + \delta X \, ,
\label{eqn:ReCha1a}  \\
\omega_{L}\tilde{\chi}^+_i & \longrightarrow & \left(\delta_{ij} +
  \frac{1}{2}\left (\delta Z^L\right )_{ij}\right) 
\omega_{L}\tilde{\chi}^+_j\, ,
\nonumber \\
\omega_{R}\tilde{\chi}^+_i& \longrightarrow &\left(\delta_{ij} +
  \frac{1}{2}\left (\delta Z^R\right )^{\ast}_{ij}\right) 
\omega_{R}\tilde{\chi}^+_j\, ,
\label{eqn:ReCha1b} 
\end{eqnarray}
where $\omega_{L,R}=(1 \mp \gamma_5)/2$. Each element of $\delta X$ is the
counterterm for the corresponding entry in $X$; in particular, its diagonal
entries are the counterterms $\delta M_2, \delta \mu$. As for the fermionic
fields of the SM, we need to introduce independent field renormalization
constants for the left- and right-handed components of $\tilde \chi_i^+$.
These constants $\delta Z^L$ and $\delta Z^R$ are general
$2\times 2$ matrices.

Similarly to the chargino case, we introduce renormalization constants for the
neutralino mass matrix $Y$ and for the physical four-component (Majorana)
neutralino fields $\tilde{\chi}^0_i (i =1, 2, 3, 4)$ \cite{ReNeu}:
\begin{eqnarray}
Y &\longrightarrow & Y + \delta Y \, ,
\label{eqn:ReNeu1a}  \\
\omega_{L}\tilde{\chi}^0_i &=&\left(\delta_{ij} + \frac{1}{2}\left
    (\delta Z^0\right )_{ij}\right) \omega_{L}\tilde{\chi}^0_j\, ,
\nonumber \\
\omega_{R}\tilde{\chi}^0_i& =&\left(\delta_{ij} + \frac{1}{2}\left
    (\delta Z^0\right )^{\ast}_{ij}\right)
\omega_{R}\tilde{\chi}^0_j\, .
\label{eqn:ReNeu1b}
\end{eqnarray}
Here the elements of $\delta Y$ are the counterterms for the corresponding
entries in $Y$; in particular, the diagonal $2\times 2$ blocks contain the
counterterms $\delta M_1, \, \delta M_2$, and $\delta \mu$. The field
renormalization constant $\delta Z^0$ is a general complex $4\times 4$
matrix. Note that the Majorana condition $\tilde \chi_i^0 = \left( \tilde
  \chi_i^0 \right)^C$ implies that the left- and right-handed components of
$\tilde \chi_1^0$ do {\em not} renormalize independently, as shown in
(\ref{eqn:ReNeu1b}).

In the on-shell renormalization scheme for the charginos/neutralinos
\cite{ReNeu} the counterterms $\delta M_2, \delta \mu$, and $\delta M_1$ are
determined by requiring that the masses of $\tilde \chi_1^+, \, \tilde
\chi_2^+$ and $\tilde \chi_1^0$, which are defined as the poles of the
corresponding propagators, are the same as at tree-level. We have slightly
modified this prescription, keeping $m_{\tilde \chi_1^0}, \, m_{\tilde
  \chi_2^0}$ and $m_{\tilde \chi_2^\pm}$ fixed, since $m_{\tilde \chi_2^0}$ is
obviously more important for our analysis than $m_{\tilde \chi_1^\pm}$. The
diagonal entries of the field renormalization constants are fixed by the
condition that the corresponding renormalized propagator has unit residue.
Furthermore, the renormalized one-particle irreducible two-point functions
should be diagonal for on-shell external particles, which fixes the
off-diagonal entries of the field renormalization constants. We note here
that in this scheme the masses of the heavier neutralinos $\tilde \chi^0_{3,
  \, 4}$ and lighter chargino $\tilde \chi_1^\pm$ do differ from their input
(tree-level) values after one-loop corrections have been included. However,
these shifts violate the electroweak $SU(2)$ symmetry, and are therefore
usually quite small at least for gaugino-like states.\footnote{In the
  presence of strong $L-R$ mixing in the stop sector, the masses of the
  higgsino-like neutralinos can still be shifted by several GeV
  \cite{ReNeu, higgsino}.} In most of mSUGRA parameter space (as well as in
many other scenarios) the gaugino-like states are lighter than the
higgsino-like ones. Moreover, gaugino-like states are usually produced more
copiously in the decays of squarks and gluinos.

\subsection{Renormalization of the Sfermion sector}\label{ReSle}

In general the superpartners $\tilde{f}_L, \, \tilde{f}_R$ of the fermions
$f_L, \, f_R$ mix to form the sfermion mass eigenstates $\tilde{f}_s \ (s = 1,
2)$. The MSSM does not contain right-handed neutrino superfields, hence there 
is no $L-R$ mixing in the sneutrino sector. We assume that sfermions of 
different flavors do not mix. 
We renormalize the sfermion mass matrices 
$M_{\tilde{f}}$ and the sfermion fields $\tilde{f}_s \ (s = 1, 2)$ via
\begin{eqnarray}
M_{\tilde{f}} &\longrightarrow & M_{\tilde{f}} + \delta M_{\tilde{f}}\, , 
\label{eqn:ReSle1a}  \\
\tilde{f}_s &\longrightarrow & \left(\delta_{st} + \frac{1}{2}\left
    (\delta Z_{\tilde{f}}\right )_{st}\right)\tilde{f}_t\, . 
\label{eqn:ReSle1b}
\end{eqnarray}
The elements of the matrices $\delta M_{\tilde{f}}$ are the counterterms for 
the corresponding entries in $M_{\tilde{f}}$. The 
field renormalization constants $\delta Z_{\tilde{f}}$ are general
$2\times 2$ matrices. For the sneutrinos, their masses, their counterterms, and
the field renormalization constants are simple numbers rather than matrices.

We follow Ref.~\cite{ReSle} and renormalize the sfermion sector via the 
on-shell scheme. For every generation of the squarks, the independent 
parameters are the soft-breaking mass parameters 
$M_{\tilde u_L}^2 = M_{\tilde d_L}^2 \equiv M_{\tilde q_L}^2, \, 
M_{\tilde{u}_R}^2,\, M_{\tilde{d}_R}^2$ and the scalar 
trilinear coupling parameters $A_u$,\,  $A_d$. In order to fix their 
counterterms, one can renormalize two up-squarks and the lighter 
down-type squark via the on-shell renormalization scheme. 
It requires that the renormalized masses of the two up-type squarks and the 
lighter down-type squark are equal to their physical (input) masses, and
that the renormalized two-point function is diagonal for on-shell external
particles. These on-shell conditions determine the counterterms 
$\delta M_{\tilde q_L}^2, \, \delta M_{\tilde{u}_R}^2, \, 
\delta M_{\tilde{d}_R}^2$ as well as $\delta A_u$,\,  $\delta A_d$ and the 
off-diagonal entries of the field renormalization constant, under the 
assumption $\delta Z_{\tilde{q}_{12}} = \delta Z_{\tilde{q}_{21}}$. The 
independent parameters for the sleptons are $M_{\tilde{l}_L}^2$, 
$M_{\tilde{l}_R}^2$ and $A_l$. In analogy to the squarks, Ref. \cite{ReSle} 
renormalized the sleptons via imposing the on-shell renormlaization conditions 
on the sneutrino and the lighter charged slepton.
In a slight deviation from Ref. \cite{ReSle} we fix both
charged slepton masses at their tree-level values. In general there will
therefore be a shift of the mass of the sneutrino when one-loop corrections
are included. However, since this shift again vanishes in the limit of exact
electroweak gauge symmetry it is numerically very small
\cite{ReSle, yamada_slep}. Similarly to the chargino/neutralino case, the
diagonal entries of $\delta Z_{\tilde{f}}$ and the sneutrino field
renormalization constant $\delta Z_{\tilde{\nu}_l}$ are fixed by the
requirement that the corresponding propagator has unit residue.
Besides the soft-breaking sfermion mass parameters 
and the scalar trilinear coupling parameters, 
the sfermion mass matrices $M_{\tilde{f}}$ also depend on $\mu$,
whose renormalization was discussed above, as well as on $\tan\beta$,
$m_Z$, $\theta_W$, the electric charge $e$ and the
charged fermion masses $m_f$, whose renormalization will be discussed below.

\subsection{Renormalization of the neutral Higgs sector}

The renormalization of the Higgs sector in the CP-violating MSSM has been
described in Ref. \cite{ReHiggs}; here we limit ourself to the neutral Higgs 
sector of the CP-conserving MSSM, using a mixture of on-shell and 
$\overline{DR}$ renormalization.

The independent parameters in the Higgs sector are chosen to be the tadpoles
$T_{h^0}, \, T_{H^0}$ of the physical CP-even scalars $h^0$ and $H^0$, which
vanish at tree-level, the mass of the physical neutral CP-odd Higgs
boson $m_A^2$, and the ratio of VEVs $\tan\beta$ introduced above. In addition 
the counterterms from the renormalization of the weak gauge boson sector,
described below, enter here.

In the neutral CP-odd Higgs boson sector, the mass matrix $M_{\chi^0}$ and
the fields $A^0, \, G^0$ are renormalized via
\begin{eqnarray}
M_{\chi^0}&\to & M_{\chi^0} + \delta M_{\chi^0}\, ,\\
\left(\begin{array}{c} A^0 \\ G^0 \end{array}\right) 
&\to &  \left (\begin{array}{cc}
1 + \frac{1}{2}\delta Z_{AA} & \frac{1}{2}\delta Z_{AG}\\
\frac{1}{2}\delta Z_{GA}& 1 + \frac{1}{2}\delta Z_{GG}
\end{array}\right)
\left(\begin{array}{c} A^0 \\ G^0 \end{array}\right)\, .
\end{eqnarray}
Similarly, the neutral CP-even Higgs boson sector is renormalized as follows:
\begin{eqnarray}
M_{\phi^0}&\to & M_{\phi^0} + \delta M_{\phi^0}\, ,
\\
\left(\begin{array}{c} h^0 \\ H^0 \end{array}\right) &\to &   
\left (\begin{array}{cc}
1 + \frac{1}{2}\delta Z_{hh} & \frac{1}{2}\delta Z_{hH}\\
\frac{1}{2}\delta Z_{Hh}& 1 + \frac{1}{2}\delta Z_{HH}
\end{array}\right)
\left(\begin{array}{c} h^0 \\ H^0 \end{array}\right)\, ,
\end{eqnarray}
$M_{\phi^0}$ is the mass matrix of the CP-even Higgs bosons and the matrices
$\delta M_{\chi^0}$, $\delta M_{\phi^0}$ contain the counterterms $\delta
T_{h^0}, \delta T_{H^0}$, $\delta m_A^2$, and $\delta \tan\beta$.

The tadpole counterterms are fixed by the requirement that the renormalized
tadpoles vanish. The counterterm $\delta m_{A}^2$ is determined by on-shell
renormalization of the neutral CP-odd Higgs boson $A^0$. In this paper, we
fix the field renormalization constants in the Higgs sector as well as $\delta
\tan\beta$ in the $\overline{DR}$ scheme, which means that the counterterms
only contain UV-divergent parts (plus some process-independent numerical 
constants).\footnote{In the numerical examples discussed below, Higgs exchange
  contributions are negligible. However, they will be significant if $\tilde
  \chi_2^0 \rightarrow h^0 \tilde \chi_1^0$ decays are open, and/or at
  high $\tan\beta$, where the Yukawa couplings to charged leptons and charge
  1/3 quarks are enhanced.} In case of $\tan\beta$, this implies
\begin{eqnarray}
\frac{\delta\tan\beta}{\tan\beta}& = &
\frac{1}{2m_Z\sin\beta\cos\beta} \bigl[ \mathrm{Im} \Sigma_{A^0Z}(m_A^2)
\bigr]^{div}\, . 
 \end{eqnarray}
 Other ways to renormalize $\tan\beta$ are discussed in Refs.~\cite{RetanB, 
Dominik, FrankHiggs}.

\subsection{Renormalization of the SM-like sector}

The final piece of the Lagrangian we have to renormalize contains terms also
present in the SM. Here we follow Ref.~\cite{SMDenner}. The relevant
parameters are the electric charge $e$, the charged fermion masses $m_f$, and
the masses of the $W, Z$ bosons. They are renormalized as follows:
\begin{eqnarray}
e &\rightarrow& \left (1 + \delta Z_e\right)e\, ,\\
m_f &\rightarrow& m_f + \delta m_f\, ,\\
m_{W,Z}^2 &\rightarrow & m_{W,Z}^2 + \delta  m_{W,Z}^2\, .
\end{eqnarray}
The wave function renormalization of the fermion and neutral vector boson
fields is described by
\begin{eqnarray}
f_i^L &\rightarrow &\left( \delta_{ij} + \frac{1}{2}\delta Z_{ij}^{f,L}\right)f_j^L\,
,\\ 
f_i^R &\rightarrow &\left( \delta_{ij} + \frac{1}{2}\delta Z_{ij}^{f,R}\right)f_j^R\,
,\\ 
\left(\begin{array}{c} Z \\ A \end{array}\right)
&\rightarrow & \left(\begin{array}{cc}
1+ \frac{1}{2}\delta Z_{ZZ} & \frac{1}{2}\delta Z_{ZA}\\
\frac{1}{2} \delta Z_{AZ} & 1+ \frac{1}{2}\delta Z_{AA}
\end{array}\right)\left(\begin{array}{c} Z \\ A \end{array}\right)\, .
\end{eqnarray}
The renormalization constants above are again fixed by the on-shell
conditions \cite{SMDenner}. The on-shell definition of the weak mixing angle
$\theta_W(s_W =\sin\theta_W, c_W = \cos\theta_W)$ is \cite{sirlin}
\begin{eqnarray}
s_W^2 &= &1- \frac{m_W^2}{m_Z^2}\, .
\end{eqnarray}
Hence its counterterm is directly related to the counterterms of the gauge
boson masses,
\begin{eqnarray}
\frac{\delta s_W}{s_W} &= & - \frac{1}{2} \frac {c_W^2} {s_W^2}
\left( \frac {\delta m_W^2} {m_W^2} - \frac {\delta m_Z^2} {m_Z^2} \right)\, .
\end{eqnarray}
This completes our discussion of the renormalization conditions. We are now
ready to discuss the calculation of the $\tilde \chi_2^0$ decay width.

\section{Tree-level calculations for $\tilde \chi_2^0 \to \tilde \chi_
1^0 l^- l^+$}\label{tree}

\begin{figure}[b!]
\begin{center}
\begin{tabular}{cccc}
\includegraphics[width=0.20\linewidth]{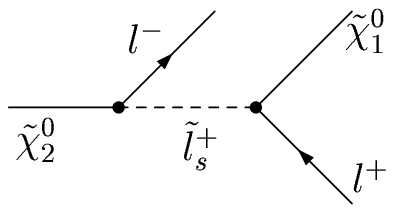} &
\includegraphics[width=0.20\linewidth]{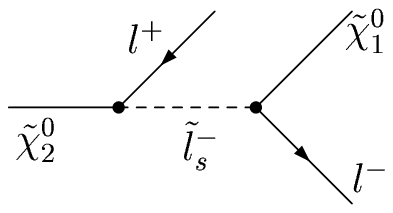}& 
\includegraphics[width=0.20\linewidth]{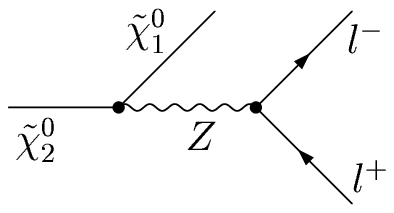} &
\includegraphics[width=0.20\linewidth]{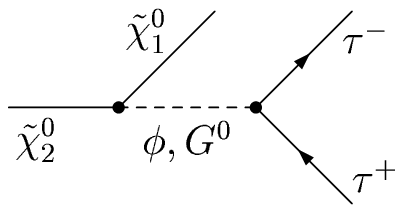} \\
(a) & (b) & (c) & (d)
\end{tabular}
\end{center}
\caption{The Born Feynman diagrams for $\tilde{\chi}_2^0 \longrightarrow
  \tilde{\chi}_1^0 l^- l^+ (l = e, \mu, \tau)$. $s = 1,2$ labels the slepton
  mass eigenstates, $\phi$ denotes the MSSM neutral Higgs boson $h^0, H^0, 
 A^0$, and the neutral Goldstone boson $G^0$ which appears only together with 
the $Z$ boson in using a non-unitary gauge. Since the Yukawa coupling 
$\phi l^- l^+$ is proportional to the lepton mass, the Higgs intermediate 
states are neglected when $l = e$ and  $\mu$. 
\label{treediagram}}
\end{figure}

The Born Feynman diagrams for $\tilde{\chi}_2^0 \longrightarrow
\tilde{\chi}_1^0 l^- l^+ (l = e, \mu, \tau)$ are displayed in
Fig.~\ref{treediagram}. The propagators of the diagrams (a) and (b) have the
structure as
\begin{equation}
\frac{1}{k^2 - m_{\tilde{l}_s}^2}\, ,
\label{eqn:tree1}
\end{equation}  
where $k$ and $m_{\tilde{l}_s}$ denote the 4-momentum of the propagator and
the slepton mass, respectively.  
If the two-body decays $\tilde{\chi}_2^0\rightarrow \tilde{l}^\pm_1 l^\mp 
\rightarrow \tilde{\chi}_1^0 l^- l^+$ are kinematically allowed, i.e. the 
sleptons $\tilde{l}_1$ can be on shell at some points in the phase space, 
a finite width of $\tilde{l}_1$ is necessary. 
It arises from the imaginary part of the slepton self-energy.
A finite width is introduced via Dyson summation,
\begin{eqnarray}
\frac{i}{k^2 -  m_{\tilde{l}_1}^2}+ \frac{i}{k^2 - m_{\tilde{l}_1}^2}
 i \hat{\Sigma}(k^2) \frac {i} {k^2 - m_{\tilde{l}_1}^2 }
 + \cdots & = & \frac{i}{k^2 -  m_{\tilde{l}_1}^2+ \hat{\Sigma}(k^2)}\, ,
\label{eqn:completetree1a}
 \end{eqnarray}
 where $\hat{\Sigma}(k^2)$ is the renormalized $\tilde{l}_1$
 self-energy.

A gauge invariant matrix element is obtained
by a Laurent expansion around the complex pole~\cite{douple-pole};
in on-shell renormalization
\begin{eqnarray}
\frac{1}{k^2 -  m_{\tilde{l}_1}^2+ \hat{\Sigma}(k^2)}& \simeq &  
\frac{1}{k^2 -  m_p^2} 
\left(1 - \frac{{\rm Re}\hat{\Sigma}(k^2)}{k^2 -
    m_p^2}\right)\, , 
\label{eqn:tree2}
\end{eqnarray}
were $m_p^2$ denotes the position of the complex pole in 
(\ref{eqn:completetree1a}). It is obtained as the solution of
\begin{eqnarray}
m_p^2 - m_{\tilde{l}_1}^2 +\hat{\Sigma}(m_p^2)=0\, .
\label{eqn:complexpole-eq}
\end{eqnarray}

For the tree-level amplitude the complex pole $m_p^2$ is calculated at 
one-loop level. Its explicit expression is 
\begin{eqnarray}
m_p^2 = m_{\tilde{l}_1}^2 -  i {m_{\tilde{l}_1}}
\Gamma_{\tilde{l}_1}^{\rm tree}\, ,
\end{eqnarray}
where we have employed on-shell renormalization as in 
Sec.~\ref{ReSle}.
$\Gamma_{\tilde{l}_1}^{\rm tree}$ is the tree-level decay width of
$\tilde{l}_1$ and $m_{\tilde{l}_1} \Gamma_{\tilde{l}_1}^{\rm tree}$ is the
imaginary part of the slepton self-energy $\Sigma(m_{\tilde{l}_{1}}^2)$. The 
first factor in (\ref{eqn:tree2}) is nothing but the Breit-Wigner propagator.
Since the second term in the parentheses in (\ref{eqn:tree2}) is at one-loop 
level, we do not need it in the tree-level calculation.
Therefore, the gauge invariant tree-level amplitude $M_{\rm tree}$ for the 
decays $\tilde\chi_2^0 \rightarrow \tilde \chi_1^0 l^- l^+$
can be written as
\begin{eqnarray}\label{eqn:propM0}
M_{\rm tree}& = & 
\frac{V^{\rm tree}_{\tilde{\chi}_2^0\tilde{l}_1^{\pm}l^{\mp}}(k^2) 
V^{\rm tree}_{\tilde{l}_1^{\pm}\tilde{\chi}_1^0 l^{\pm}}(k^2)}{k^2 -
m_{\tilde{l}_1}^2 + i m_{\tilde{l}_1} \Gamma_{\tilde{l}_1}^{\rm tree}} +
B(k^2)\, , 
\end{eqnarray}
where $V^{\rm tree}_{\tilde{\chi}_2^0\tilde{l}_1^{\pm}l^{\mp}}$ and 
$V^{\rm tree}_{\tilde{l}_1^{\pm}\tilde{\chi}_1^0 l^{\pm}}$ 
represent the $\tilde{\chi}_2^0\tilde{l}_1^{\pm}l^{\mp}$ and $\tilde{l}_1^{\pm}
\tilde{\chi}_1^0 l^{\pm}$ vertices, respectively, and $B(k^2)$ denotes the 
non-resonant part of the matrix element, i.e.
the matrix element of diagrams (a) and (b) for $s=2$ and
diagrams (c) and (d) in Fig. \ref{treediagram}.

The non-resonant part is much smaller than the resonant one (diagrams (a) and 
(b) for $s=1$ in Fig. \ref{treediagram}), hence it can be neglected 
approximately. We can then compute the relevant partial widths in the 
single-pole approximation,
where the decays $\tilde{\chi}_2^0 \rightarrow \tilde{\chi}_1^0 l^- l^+$ are
treated as the production and decay of the sleptons $\tilde{l}_1$,
\begin{eqnarray}
\Gamma( \tilde{\chi}_2^0 \rightarrow  \tilde{\chi}_1^0 l^- l^+)_{\rm tree}  
& \simeq &\Gamma( \tilde{\chi}_2^0 \rightarrow \tilde{l}_1^{\pm} 
 l^{\mp})_{\rm tree} Br(\tilde{l}_1^{\pm}\rightarrow \tilde{\chi}_1^0
 l^{\pm})_{\rm tree}\, , 
\label{eqn:tree4}
\end{eqnarray}
where the branching ratio of the decay $\tilde{l}_1^{\pm}\rightarrow 
\tilde{\chi}_1^0  l^{\pm}$ is defined by 
\begin{eqnarray}
Br(\tilde{l}_1^{\pm}\rightarrow \tilde{\chi}_1^0 l^{\pm})_{\rm tree} & =
&\frac{ {\Gamma(\tilde{l}_1^{\pm} \rightarrow  
\tilde{\chi}_1^0 l^{\pm})}_{\rm tree}} {\Gamma_{\tilde{l}_1}^{\rm tree}}\, .
\label{eqn:tree5}
\end{eqnarray}
The feature that the single-pole approximation reproduces the $\tilde \chi_2^0$
partial width can be seen from the identity
\begin{equation} \label{propid}
\int_{-\infty}^{\infty} dk^2 \frac {1} {\left| k^2 - m_{\tilde l_1}^2 + i
    \Gamma_{\tilde l_1}^{\rm tree} m_{\tilde l_1} \right|^2} = 
\frac {\pi} {m_{\tilde l_1} \Gamma_{\tilde l_1}^{\rm tree}}\, .
\end{equation}
If $\Gamma_{\tilde l_1}^{\rm tree} \ll m_{\tilde l_1}$ the integral in 
(\ref{propid}) will be dominated by the regions of $k^2$ close to 
$m^2_{\tilde l_1}$, i.e. only a narrow range of $k^2$ will contribute 
significantly. Moreover, the squared 
$V_{\tilde{l}_1^{\pm}\tilde{\chi}_1^0 l^{\pm}}^{\rm tree}$
is proportional to the $\tilde l_1^\pm \rightarrow \tilde \chi_1^0 l^\pm$ 
partial width; together with the factor $1/\Gamma_{\tilde l_1}^{\rm tree}$ from
(\ref{propid}) this reproduces the factor $Br(\tilde l_1^\pm \rightarrow 
\tilde \chi_1^0 l^\pm$) in (\ref{eqn:tree4}).

Here we concentrate on scenarios where only the lighter charged
sleptons $\tilde l_1$ can be produced in $\tilde \chi_2^0$ decays; scenarios
where sneutrinos and/or heavier charged sleptons can also be produced in
two-body decays of $\tilde \chi_2^0$ can be treated analogously.

\section{One-loop calculations for $\tilde \chi_2^0 \to \tilde \chi_
1^0 l^- l^+$}\label{oneloop}

The single-pole approximation can also be used at the one-loop level;
however, we will first describe the complete calculation.

\subsection{Complete one-loop calculation}\label{complete}

\subsubsection{Virtual corrections}

In general the virtual one-loop corrections to three-body decays can be
classified as {\em self-energy contributions}, {\em vertex contributions} and
{\em box contributions}. The first two classes are UV finite only after adding
the contributions from the counterterms that originate from the
renormalization of the MSSM, as discussed in Sec.~\ref{ReMSSM}; the box
diagrams are by themselves UV finite. The MSSM Feynman rules, as well as the
resulting counterterms, are implemented in the {\sc FeynArts} package of
computer programs \cite{FeynArts}, which allows an automated generation of the 
Feynman diagrams. The matrix element and the one-loop integrals are calculated 
with the help of the packages {\sc FormCalc} and {\sc LoopTools} 
\cite{FormCLT}, respectively.

Similarly to the tree-level case, diagrams with a slepton $\tilde l_1$
propagator have singularities when $\tilde l_1$ can be on shell.
We remove the singularities by introducing a finite width of $\tilde l_1$ as 
in (\ref{eqn:completetree1a}). Following the strategy in Sec.~\ref{tree}, one 
can obtain a gauge invariant matrix element at one-loop level.  
In order to obtain $\mathcal{O}(\alpha)$ accuracy near the $\tilde{l}_1$ 
resonance, one needs to calculate the complex pole $m_p^2$ to two-loop level 
\cite{douple-pole},
\begin{eqnarray}
m_p^2 = m_{\tilde{l}_1}^2 -  i m_{\tilde{l}_1} 
\Gamma_{\tilde{l}_1}^{\rm 1-loop}\, ,
\end{eqnarray}
where we have applied the on-shell renormalization scheme at two-loop level, 
and $\Gamma_{\tilde{l}_1}^{\rm 1-loop}$ denotes the one-loop-level width 
of $\tilde l_1$.
Then the gauge invariant matrix element at one-loop level can be written as
\begin{eqnarray}
M_{\rm tree} + M_{\rm virt} & = &\frac{A(k^2)}
{k^2 -  m_{\tilde{l}_1}^2 + i m_{\tilde{l}_1}
\Gamma_{\tilde{l}_1}^{\rm 1-loop}} + C(k^2)\, , 
\label{eqn:complete3a}
\end{eqnarray}
where $ C(k^2)$ denotes the non-resonant part of the matrix element,
the residue $A(k^2)$ can be expressed as
\begin{eqnarray}
A(k^2) &= & V^{\rm tree}_{\tilde{\chi}_2^0\tilde{l}_1^{\pm}l^{\mp}}(k^2)
 V^{\rm tree}_{\tilde{l}_1^{\pm}\tilde{\chi}_1^0 l^{\pm}}(k^2)
\left(1 - \frac{{\rm Re} \hat{\Sigma}(k^2)}{k^2 -
m_{\tilde{l}_1}^2}\right)+ \, \nonumber \\
&&{} V^{\rm tree}_{\tilde{\chi}_2^0\tilde{l}_1^{\pm}l^{\mp}}(k^2)
\hat V^{\rm 1-loop}_{\tilde{l}_1^{\pm}\tilde{\chi}_1^0 l^{\pm}}(k^2)
+\hat V^{\rm 1-loop}_{\tilde{\chi}_2^0\tilde{l}_1^{\pm}l^{\mp}}(k^2)
 V^{\rm tree}_{\tilde{l}_1^{\pm}\tilde{\chi}_1^0 l^{\pm}}(k^2)\, ,
\end{eqnarray}
where $\hat V^{\rm 1-loop}_{\tilde{\chi}_2^0\tilde{l}_1^{\pm}l^{\mp}}$ and 
$\hat V^{\rm 1-loop}_{\tilde{l}_1^{\pm}\tilde{\chi}_1^0 l^{\pm}}$ 
represent the renormalized one-loop corrections to the 
$\tilde{\chi}_2^0\tilde{l}_1^{\pm}l^{\mp}$ and 
$\tilde{l}_1^{\pm}\tilde{\chi}_1^0 l^{\pm}$ 
vertices, respectively.

\begin{figure}[t!]
\begin{center}
\begin{tabular}{ll}
\includegraphics[width=0.3\linewidth]{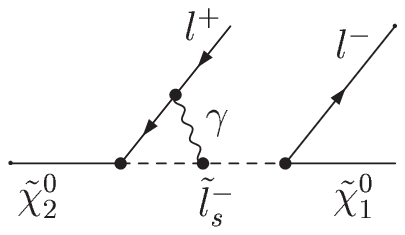} &
\includegraphics[width=0.3\linewidth]{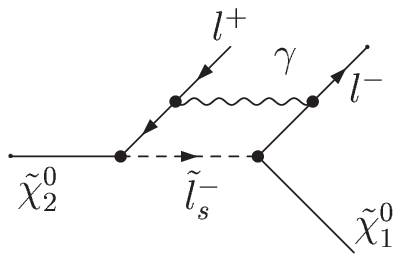} 
\end{tabular}
\end{center}
\caption{Examples for the virtual photonic corrections\label{photon}}
\end{figure}

Feynman diagrams like those shown in Fig.~\ref{photon} also give
singularities when the sleptons $\tilde{l}_1$ are on shell. The left (vertex)
diagram has an infrared (IR) divergence if a real slepton mass is used in 
kinematic configurations where the slepton can be on shell. The right (box) 
diagram has a divergence which can be understood as being due to re-scattering 
of the two charged leptons in the final state, which persists for large photon
virtualities. 
One should therefore use complex slepton masses, 
\begin{eqnarray}
\frac{1}{k^2 - m_{\tilde{l}_1}^2}& \longrightarrow & \frac{1}{k^2 -
  m_{\tilde{l}_1}^2 + i m_{\tilde{l}_1}\Gamma_{\tilde{l}_1}^{\rm 1-loop}}
\label{eqn:complete3}
\end{eqnarray}
in the one-loop integrals from these diagrams.
This gives rise to a large QED logarithm 
$\log \left ( m_{\tilde{l}_1}/\Gamma_{\tilde{l}_1}^{\rm 1-loop}\right )$.
Furthermore, the box diagram shown in  Fig.~\ref{photon} has the property that 
the virtual photon is attached to external on-shell charged particles. 
This results in IR divergences, which we regularized by introducing a
fictitious photon mass $\lambda$. The IR divergences cancel after we add
contributions from real photon bremsstrahlung, which will be discussed in 
Sec.~\ref{real}. The one-loop integrals with complex masses can be calculated 
with the help of {\sc LoopTools}. The analytical expressions for scalar 
three-point and four-point functions with real arguments can be found in 
Refs.~\cite{SMDenner, soft, loopfunction}. 
We generalized these to allow for complex arguments. 
Note that in our calculation the masses of the light leptons, 
i.e. $m_l~(l = e,\mu)$, are neglected except when they appear in the one-loop 
integrals, while the $\tau$ mass $m_\tau$ is kept everywhere.

\subsubsection{Real photon bremsstrahlung}\label{real}

In order to cancel the IR divergences in the virtual corrections, we have to 
add contributions from real photon bremsstrahlung,
which contain on-shell
propagators of {\em stable} particles in the limit where the scalar product of
the 4-momenta of the photon and the emitting charged particle vanishes.

This always happens, regardless of the mass of the emitting particle, if the
photon energy $E_\gamma$ is very small. The ``soft photon bremsstrahlung''
contribution is defined via the condition $E_{\gamma}\le \Delta E$; here the
cutoff parameter $\Delta E$ should be small compared to the relevant physical
energy scale (e.g. the resolution of the experimental apparatus). This
IR-divergent contribution is sufficient to cancel the IR divergences from
the virtual corrections. Since the energy of the emitted soft photon is by
definition very small, this emission essentially does not change the momenta
of the other final state particles. 
This contribution is therefore described as a convolution of the differential 
tree--level decay width with a universal factor.
Explicit expressions can be found in Refs. \cite{SMDenner, soft}.

Real emission contributions with $E_\gamma > \Delta E$ are called ``hard
photon bremsstrahlung''. Altogether,
\begin{eqnarray} \label{brems1}
\Gamma_{\rm brems}& = & \Gamma_{\rm soft}(\Delta E) + \Gamma_{\rm hard}(\Delta
E)\, .  
\end{eqnarray}
The dependence on the largely arbitrary parameter $\Delta E$ cancels after
summing soft and hard contributions, provided it is sufficiently small. In
the limit of vanishing mass of the emitting particle, the hard photon
bremsstrahlung contribution also contains a divergence, if the momenta of the
photon and the emitting particle are collinear. Since there are no massless
charged particles in Nature, this is not a real divergence; in our case, it is
regularized by the masses of the leptons in the final state. However, since
the lepton masses, i.e. $m_e$ and $m_\mu$, are very small, it is very
difficult to get stable numerical results from a direct numerical evaluation
of hard photon bremsstrahlung, e.g. using Monte Carlo integration. 

This can be overcome by dividing hard photon bremsstrahlung into a collinear
part, where the angle between the photon and the radiating particle is smaller
than a very small angle $\Delta \theta$, and the complementary non-collinear
part,
\begin{eqnarray} \label{brems2}
\Gamma_{\rm hard}(\Delta E)& = & \Gamma_{\rm coll}(\Delta E, \Delta\theta) +
\Gamma_{\rm non-coll}(\Delta E, \Delta\theta)\, . 
\end{eqnarray}
The angular cutoff $\Delta \theta$ should be so small that the emission of
photons emitted at angle $\theta < \Delta \theta$ relative to the emitting
lepton can be assumed not changing the {\em direction} of the 3-momentum of
this lepton.

If we treat a charged lepton and a collinear photon inclusively, i.e. the 
momentum of collinear photon is added to that of the emitting lepton, 
analytically the differential contribution of the collinear emissions is 
written as the differential tree-level decay width multiplied by a universal 
function \cite{collinear, axel}. This approach is for collinear-safe 
observables \cite{axel}. If one adds the soft and collinear contributions to 
the virtual corrections, all singularities ($\ln m_l$ and $\ln \lambda$) 
cancel. This is in accordance with the Kinoshita-Lee-Nauenberg theorem 
\cite{kln}. At the LHC the electron energy is determined calorimetrically:
in this case a collinear photon would hit the same cell of the calorimeter as
the lepton, so the two energies cannot be disentangled.\footnote{There is a
  minor caveat to this statement. The experimental definition of an
  ``electron'' usually requires the existence of a charged track whose energy
  - more exactly, absolute three-momentum - should not be grossly different
  from the energy measured by the calorimeter. This requirement may remove a
  few events with very hard collinear photons.} 
Hence the electron observables are defined as collinear-safe observables
in our calculation.

We also consider non-collinear-safe observables \cite{axel}, 
where the lepton and its collinear photon are not treated inclusively. 
The contribution of the collinear photon
bremsstrahlung cannot be calculated analytically.
In this case the mass singularity $\ln m_l$ cannot be canceled in the 
differential width and hence becomes visible.
Muon energies are generally measured through the curvature of their track in a
magnetic field. This measures the energy (more exactly, the 3-momentum) of
the muon {\em after} emitting the collinear photon (if any). 
In our calculation the muon observables are treated as non-collinear-safe 
observables. Finally, the
contribution from the emission of hard non-collinear photons is calculated by
using a multi-channel Monte Carlo approach \cite{mutichannel}.

The virtual photonic corrections by themselves are UV divergent\footnote{In
  principle one can define a renormalizable non-supersymmetric theory
  containing only leptons, sleptons, neutralinos and photons. However, the
  counterterms computed in this theory would be different from those of the
  full MSSM.}, hence one cannot meaningfully separate the QED corrections from
the one-loop contributions by simply selecting diagrams which contain a photon.
In the case of the light lepton ($l = e, \mu$) final states, following 
conventions of the Supersymmetric Parameter Analysis (SPA)~\cite{SPA},
we can pick out and separate potentially large QED terms from the the sum
of virtual and soft photon bremsstrahlung corrections, 
$\Gamma_{\rm virt}+ \Gamma_{\rm soft}$:
\begin{eqnarray} \label{rem}
\Gamma_{\rm virt}+ \Gamma_{\rm soft} & = & \tilde{\Gamma} + \Gamma_{\rm
  remainder}\, , 
\end{eqnarray}
where $\tilde{\Gamma}$ contains all the potentially large terms proportional
to $\log m_l$ or $\log\Delta E$, while $\Gamma_{\rm remainder}$ is IR and
UV finite and free of such large QED logarithms. The ``QED contributions" can
then be defined as follows:
\begin{eqnarray} \label{qed}
\Gamma_{\rm QED}& = &\tilde{\Gamma} + \Gamma_{\rm hard} \, .
\end{eqnarray}
Note that $\Gamma_{\rm QED}$ defined in this way does not depend on the cutoff
parameters $\Delta E$ and $\Delta \theta$. Moreover, terms proportional to 
$\log m_l$ cancel between the two contributions in $\Gamma_{\rm QED}$ in 
(\ref{qed}) (specifically, between $\tilde \Gamma$ and $\Gamma_{\rm coll}$)
in the integrated width and in the differential width for the collinear-safe 
observables. Using the definitions (\ref{rem}) and (\ref{qed}), the complete 
one-loop contribution can be written as
\begin{eqnarray} \label{llrem}
\Gamma_{\rm 1-loop} & = & \Gamma_{\rm tree} + \Gamma_{\rm virt} + \Gamma_{\rm
  brems}\, 
\nonumber \\ & = & \Gamma_{\rm tree} + \Gamma_{\rm remainder} + \Gamma_{\rm
  QED}\, . 
\end{eqnarray}

One should perform the replacement (\ref{eqn:complete3}) also in the real
photon bremsstrahlung when the sleptons $\tilde{l}_1$ can be on shell. In this
case $\tilde{\Gamma}$ contains the large QED logarithm $\log \left (
m_{\tilde{l}_1}/\Gamma_{\tilde{l}_1}^{\rm 1-loop}\right )$, besides 
$\log m_l$ and $\log\Delta E$. However, in the integrated partial width these 
terms nearly cancel after summing all contributions; more exactly, the total 
pre-factor of $\log \left (m_{\tilde l_1}/\Gamma_{\tilde l_1}^{\rm 1-loop}
\right )$ vanishes when $\Gamma_{\tilde l_1}^{\rm 1-loop} \rightarrow 0$, 
once one includes the fact
that the squared $\tilde l_1^\pm\tilde \chi_1^0 l^\mp$ vertex is $\propto
\Gamma_{\tilde l_1}$.\footnote{In the limit $\Gamma_{\tilde l_1} \rightarrow
  0$ some kinematical distributions would become singular; for example, the
  distribution in the invariant mass of the $\tilde \chi_1^0 - l^\pm$ systems
  would contain $\delta-$functions.}

When $\tau^- \tau^+$ are the final states of $\tilde \chi_2^0$ decay,  
the $\tau$ mass $m_{\tau}$ is kept everywhere. This mass is so large that
a stable numerical result can be obtained from the hard photon bremsstrahlung
even for $\Delta \theta \rightarrow 0$, i.e. we do not need to divide the hard
photon bremsstrahlung contribution into collinear and non-collinear parts.
Since we do not count terms $\propto \log m_\tau$ as a ``large logarithm'', we
follow a slightly different procedure to define the ``QED part'' of the
correction. The virtual corrections contain photonic and non-photonic
contributions,
\begin{eqnarray}
\Gamma_{\rm virt} & = & \Gamma_{\rm virt}^{\gamma} + \Gamma_{\rm
  virt}^{\rm non-\gamma}\, , 
\end{eqnarray}
both of which are UV divergent, while the sum is finite (after including all
counterterms). The photonic virtual corrections can be split into an UV-finite
part $\tilde{\Gamma}$ and an UV-divergent part $\Gamma_{\rm UV-div}^{\gamma}$,
\begin{eqnarray}
\Gamma_{\rm virt}^{\gamma} & = & \tilde{\Gamma} + \Gamma_{\rm
  UV-div}^{\gamma}\, .  
\end{eqnarray}
Here $\Gamma_{\rm UV-div}^\gamma$ contains the terms that would be subtracted
in an $\overline{DR}$ regularization of $\Gamma_{\rm virt}^\gamma$. After
this rearrangement, the virtual corrections can be written as
\begin{eqnarray}
\Gamma_{\rm virt} & = & \tilde{\Gamma} + \Gamma_{\rm UV-div}^{\gamma} +
\Gamma_{\rm virt}^{\rm non-\gamma}\, \nonumber \\ & = &  \tilde{\Gamma} +
\Gamma_{\rm remainder}\, , 
\end{eqnarray}
where $\Gamma_{\rm remainder} = \Gamma_{\rm UV-div}^{\gamma}+ \Gamma_{\rm
  virt}^{non-\gamma}$ as well as $\tilde \Gamma$ are UV finite. The ``QED
corrections" are finally defined as
\begin{eqnarray}
\Gamma_{\rm QED} & =& \tilde{\Gamma}+ \Gamma_{\rm brems}\, ,
\end{eqnarray}
where $\Gamma_{\rm brems}$ stands for the contribution from all diagrams with
real photon emission. By construction, $\Gamma_{\rm QED}$ is both UV and
IR finite. 

\subsection{One-loop calculation in the single-pole approximation}
\label{approximate} 

If $\tilde \chi_2^0 \rightarrow \tilde l_1 l$ two-body decays are allowed and
$\tilde \chi_2^0$ does not have other two-body decay modes, at one-loop
level, just like at tree level, the decays $\tilde{\chi}_2^0 \rightarrow
\tilde{\chi}_1^0 l^- l^+ $ can be approximately treated as
production and decay of the sleptons $\tilde{l}_1$,
\begin{eqnarray}
\Gamma(\tilde{\chi}_2^0 \rightarrow  \tilde{\chi}_1^0 l^- l^+)_{\rm 1-loop}  
& \simeq &\Gamma(\tilde{\chi}_2^0 \rightarrow \tilde{l}_1^{\pm}
 l^{\mp})_{\rm 1-loop} Br(\tilde{l}_1^{\pm}\rightarrow \tilde{\chi}_1^0
 l^{\pm})_{\rm 1-loop}\, ,
\end{eqnarray}
with
\begin{eqnarray}
Br(\tilde{l}_1^{\pm}\rightarrow \tilde{\chi}_1^0 l^{\pm})_{\rm 1-loop} & =
&\frac{ {\Gamma(\tilde{l}_1^{\pm} \rightarrow \tilde{\chi}_1^0 l^{\pm})}_{\rm
    1-loop}} {\Gamma_{\tilde{l}_1}^{\rm 1-loop}}\, .
\end{eqnarray}
The virtual contributions of the production and decay of the sleptons
$\tilde{l}_1$, which now only contain vertex type corrections
but no box diagrams, are again calculated with the help of the programs 
{\sc FeynArts, FormCalc} and {\sc LoopTools}. In order to obtain IR-finite
results, the real photon bremsstrahlung is added, which is again separated
into an IR-divergent soft part and an IR-finite hard part. 
For the light lepton final states $l = e, \mu$, the division of
the hard photon bremsstrahlung contribution into a collinear part, which can
be calculated analytically, and a non-collinear part, which is calculated
numerically, proceeds along the lines described in Sec.~4.1.2.
As discussed in Sec.~\ref{complete}, the UV-divergent photonic
contributions cannot be treated separately as ``QED corrections". We define
the ``QED corrections" in the same way as in the complete calculation. One
finally arrives at a total one-loop contribution which is independent of the
cutoff parameters.

\section{Total decay width of $\tilde\chi_2^0$ and the branching 
\mbox {ratios} of the decays $\tilde\chi_2^0 \to \tilde\chi_1^0 l^- l^+$}
\label{brlepton}

As discussed in Sec.~\ref{introduction}, the next-to-lightest neutralino 
$\tilde \chi_2^0$ can decay into the LSP $\tilde \chi_1^0$ and two fermions 
$f\bar f$. The leptonic final states are important because they can be 
identified at the LHC. Moreover, the endpoint of the dilepton invariant mass 
distribution is used to determine the mass relations of supersymmetric 
particles. The invisible $\tilde \chi_2^0$ decay modes, i.e. $\tilde \chi_2^0 
\to \tilde \chi_1^0 \nu_l \bar \nu_l$, do not effect the dilepton  invariant 
mass distribution. But they contribute to the total width of 
$\tilde \chi_2^0$. Since it is very difficult to identify quarks at the LHC, 
the hadronic decays $\tilde \chi_2^0 \to \tilde \chi_1^0 q \bar q$ are less 
interesting than leptonic decays. In order to obtain the total decay width of 
$\tilde \chi_2^0$, these hadronic decays must be calculated. The  total decay 
width of $\tilde \chi_2^0$ can be written as
\begin{equation}
\Gamma_{\tilde{\chi}_2^0} = \sum_{l=e,\mu,\tau} \Bigl [ \Gamma(\tilde{\chi}_2^0
\rightarrow l^- l^+ \tilde{\chi}_1^0) + \Gamma(\tilde{\chi}_2^0 \rightarrow
\nu_l \bar \nu_l \tilde{\chi}_1^0) \Bigr ] 
+ \sum_{q = u, d, c, s, b} \Gamma(\tilde{\chi}_2^0
\rightarrow q \bar q \tilde{\chi}_1^0)\, .
\end{equation}
Here we assume that the decay  $\tilde \chi_2^0 \to \tilde \chi_1^0 t \bar t$ 
is not kinematically allowed. The branching ratios of the leptonic decays  
$\tilde \chi_2^0 \to \tilde \chi_1^0 l^+ l^-$ are defined as
\begin{equation}
Br(\tilde \chi_2^0 \to \tilde \chi_1^0 l^+ l^-) = \frac{\Gamma(\tilde \chi_2^0 
\to \tilde \chi_1^0 l^+ l^-)}{\Gamma_{\tilde{\chi}_2^0}}\, .
\end{equation}

The invisible decays $\tilde \chi_2^0 \to\tilde \chi_1^0 \nu_l \bar\nu_l$ are 
calculated at tree and one-loop level. At tree level these decays can proceed 
through the exchange of sneutrinos or $Z$ bosons; cf. Fig.~1. 
Here we focus on the case where the decays $\tilde \chi_2^0 \to\tilde \chi_1^0 
\nu_l \nu_l$ are pure three-body decays. These decays are calculated similarly 
to the calculations for the leptonic three-body decays
$\tilde \chi_2^0 \to\tilde \chi_1^0 l^- l^+$.
Since none of the external particles carries electric charge in the decays 
$\tilde \chi_2^0 \to\tilde \chi_1^0\nu_l \nu_l$, there are no corrections 
involving real or virtual photons, and hence no IR divergences. 
Therefore, there are also no QED corrections in these decays.
This makes the calculation of the partial width into neutrinos
considerably simpler than for decays into charged leptons.
 
The hadronic decays of $\tilde \chi_2^0$ are calculated in order to obtain the 
total width of $\tilde \chi_2^0$. The Born Feynman diagrams for the decays 
$\tilde \chi_2^0 \to\tilde \chi_1^0 q \bar q~(q \neq t)$ are similar to those 
of Fig.~1 with intermediate squarks instead of sleptons. Here we only consider 
the case where the hadronic decays 
$\tilde \chi_2^0 \to\tilde \chi_1^0 q \bar q$ 
are pure three-body decays. Since the SUSY-QCD corrections are not considered 
in our calculations, these decays can be treated in the same way as $\tilde
\chi_2^0 \to\tilde \chi_1^0 l^- l^+$. The virtual photonic corrections of the  
hadronic decays (the diagrams are similar to the ones of the leptonic 
decays, i.e. Fig.~\ref{photon}) are IR divergent. The contributions of the 
real photon bremsstrahlung are necessary for the cancellation of the IR 
divergences. We neglect the light quark masses, i.e. $m_q (q = u, d, s)$, 
except when they appear in the one-loop integrals. In analogy to 
Sec.~\ref{real}, the contributions of the real photon bremsstrahlung are also 
splitted into an IR-divergent soft part and an IR-finite hard part. 
For the light quark final states, we separate the hard photon bremsstrahlung 
into a collinear part and a non-collinear part in order to  
obtain stable numerical results. Since quarks are detected as jets, which 
contain many photons, quark energies are always collinear-safe. 
As presented in Sec.~\ref{real}, the soft and 
collinear contributions are calculated analytically. We treat the decays with 
heavy quark final states in the same way as $\tau^-\tau^+$ final states.
The QED corrections are defined in the same way as in Sec.~\ref{real} 
since the photonic contributions are UV divergent and cannot be treated 
separately.

\section{Numerical results and discussion} \label{numerical}

We are now ready to present numerical results of our calculation. We present
results both for a scenario where $\tilde \chi_2^0$ can undergo two-body
decays, and for a scenario where no two-body decays of $\tilde \chi_2^0$ are
possible. Furthermore, we discuss both the integrated partial widths and
branching ratios of $\tilde \chi_2^0$, and the distribution of the $l^+ l^-$
invariant mass; this distribution is of great interest for future experiments,
as discussed in the Introduction. Since we always assume equal masses for
selectrons and smuons and the light lepton mass $m_l~(l = e, \mu)$ is 
neglected except when it appears in the one-loop integrals,
the integrated partial widths for the $e^+ e^- \tilde\chi_1^0$ and 
$\mu^+ \mu^- \tilde \chi_1^0$ final states are identical.  
As discussed in Sec.~\ref{real}, the dilepton invariant mass $M_{e^+e^-}$ is 
defined as collinear-safe observable, i.e. we add the momentum of a collinear 
photon to that of the emitting electron, since it is difficult to separate 
their energies at the LHC. The energies of a muon and its collinear photon 
can be disentangled easily at the LHC, hence the dilepton invariant mass 
$M_{\mu^+\mu^-}$ is defined as non-collinear-safe observable, i.e. 
the momentum of a collinear photon is not added to that of its emitter muon. 
In this case the large logarithm $\ln m_{\mu}$ can not cancel in the 
distribution, so the mass effect can be seen in the dilepton invariant mass 
distribution. One will obtain identical $M_{e^+e^-}$ and $M_{\mu^+ \mu^-}$ 
distributions if both of them are defined as collinear-safe observables.
In order to see the differences of the two treatments (adding and not adding 
the momentum of a collinear photon to the emitting lepton), we also show the 
comparison of dilepton invariant mass $M_{\mu^+\mu^-}$ and  $M_{e^+e^-}$ 
distributions.
\subsection{Numerical results for the SPS1a parameter set}
\begin{figure}[t!]
\begin{tabular}{ll}
\includegraphics[width=0.48\linewidth]{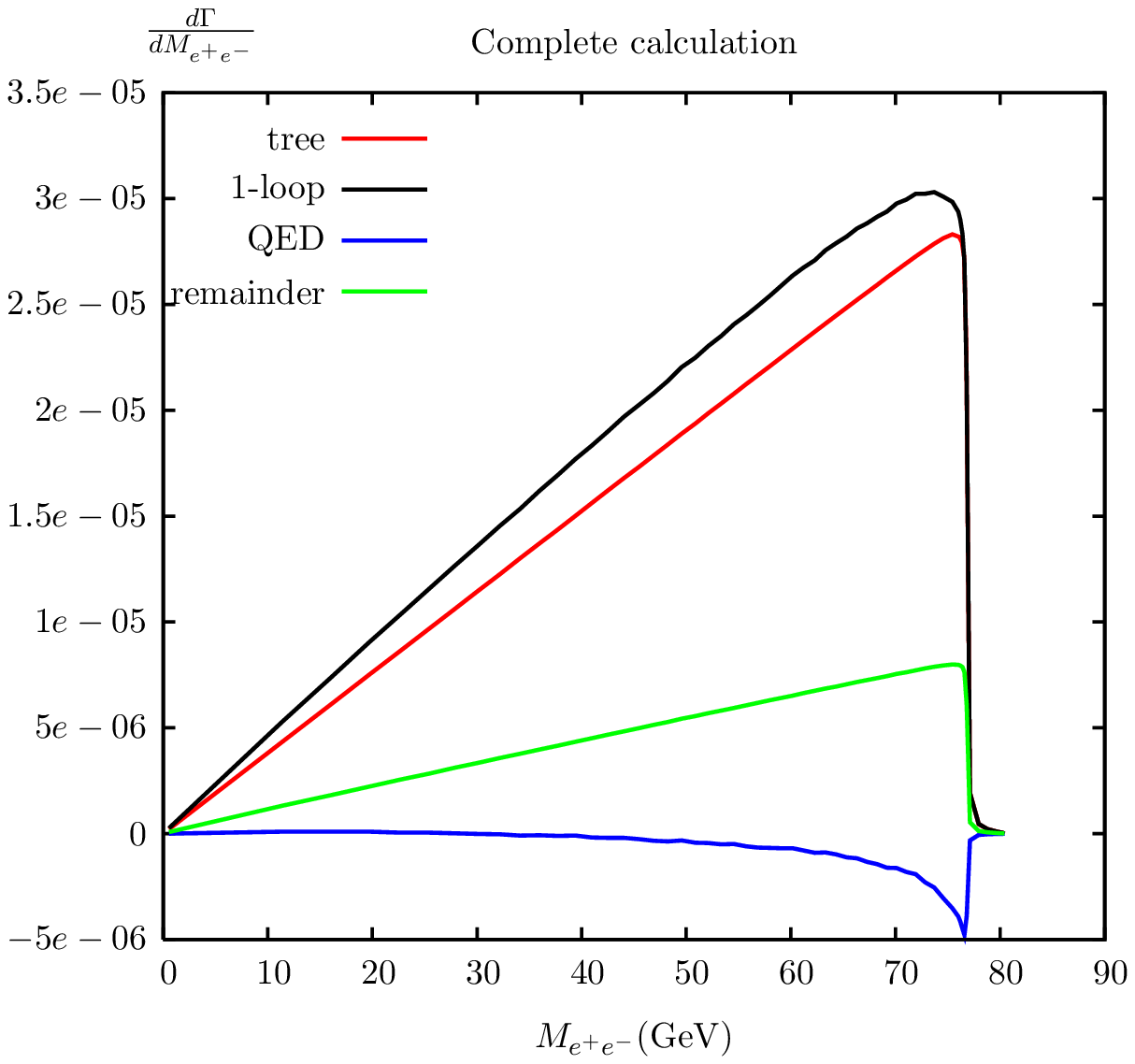}&
\includegraphics[width=0.465\linewidth]{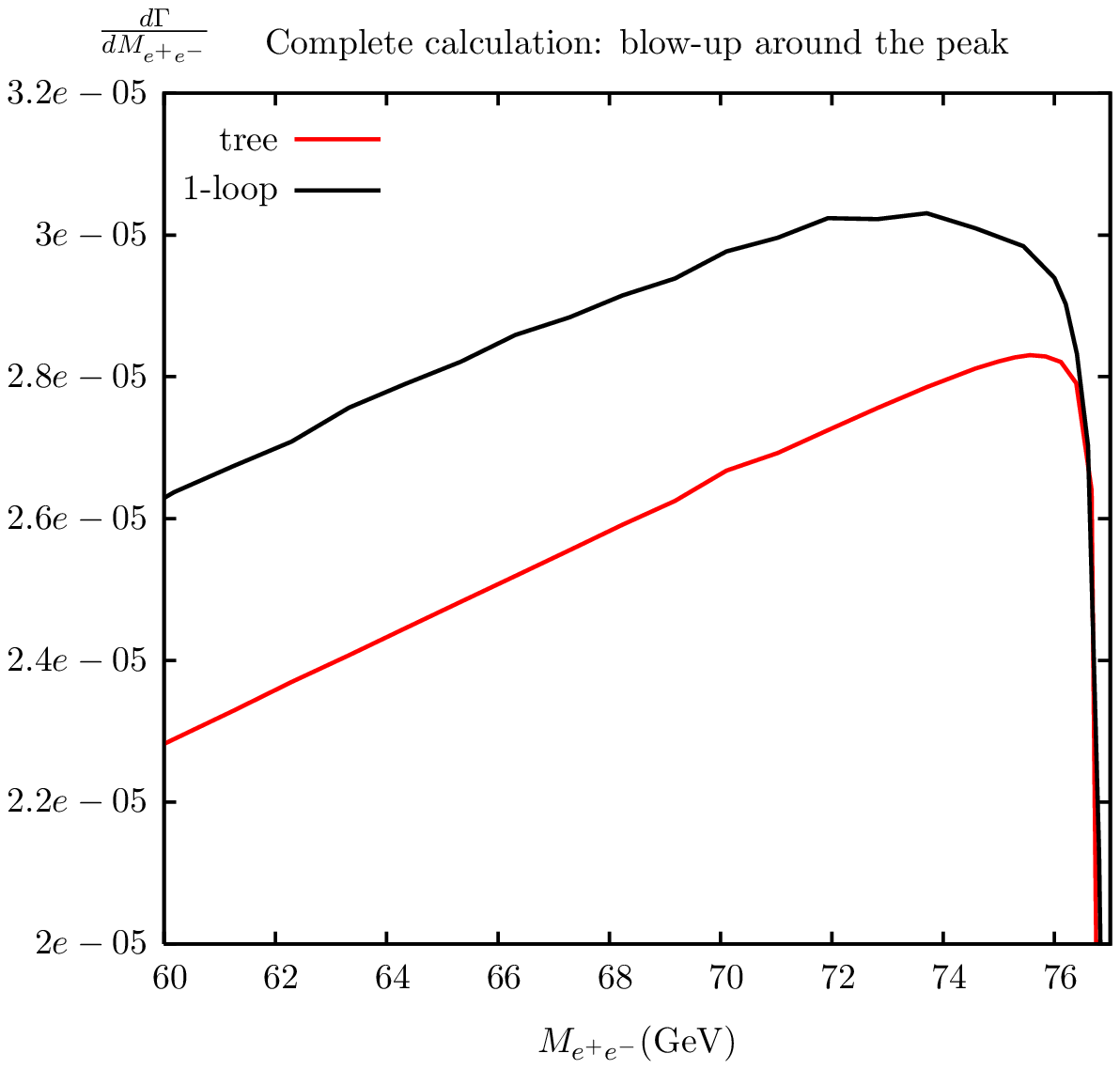}
\end{tabular}
\caption{The dilepton invariant mass $M_{e^+ e^-}$ distribution for the SPS1a
  parameter set (complete calculation). \label{SPS1aMee}}
\end{figure}

We first present results for the SPS1a 
benchmark scenario, as described in Table~1 in the
Introduction.  The dilepton invariant mass $M_{e^+ e^-}$ distribution from the
complete calculation is shown in Fig.~\ref{SPS1aMee}. In the left frame we
show not only the tree-level and total one-loop predictions, but also
the separate QED and ``remainder'' corrections, see (\ref{llrem}). We
see that the non-QED contributions are positive and quite large everywhere,
whereas the QED contribution is large and negative near the endpoint of the
distribution, but small elsewhere. In full three-body kinematics this
endpoint is simply given by $\left. M_{e^+e^-}^{\rm max}\right|_{\rm 3-body} 
= m_{\tilde \chi_2^0} -m_{\tilde \chi_1^0}$. 
However, for the SPS1a parameter set $\tilde \chi_2^0$
decays are dominated by contributions with on-shell $\tilde l_1$ in the
intermediate state. The endpoint for this two-body configuration is given by
\begin{equation}
\left. M_{e^+ e^-}^{\rm max}\right|_{\rm 2-body} =  m_{\tilde{\chi}_2^0}
  \sqrt{1- \frac{ m_{\tilde{e}^{\pm}_1}^2} {m_{\tilde{\chi}_2^0}^2}} 
\sqrt{1- \frac{m_{\tilde{\chi}_1^0}^2} {m_{\tilde{e}^{\pm}_1}^2}}  \simeq 76.8
  \ {\rm GeV} \, ,
\label{Mee_end}
\end{equation}
where the numerical value holds for the SPS1a scenario. Note that this is only
3.6 GeV below the endpoint of the three-body decays. At tree level, 
the $M_{e^+e^-}$ distribution peaks at the region which is a little below 
the endpoint of the two-body contribution. The right panel in
\begin{figure}[htb]
\begin{tabular}{ll}
\includegraphics[width=0.48\linewidth]{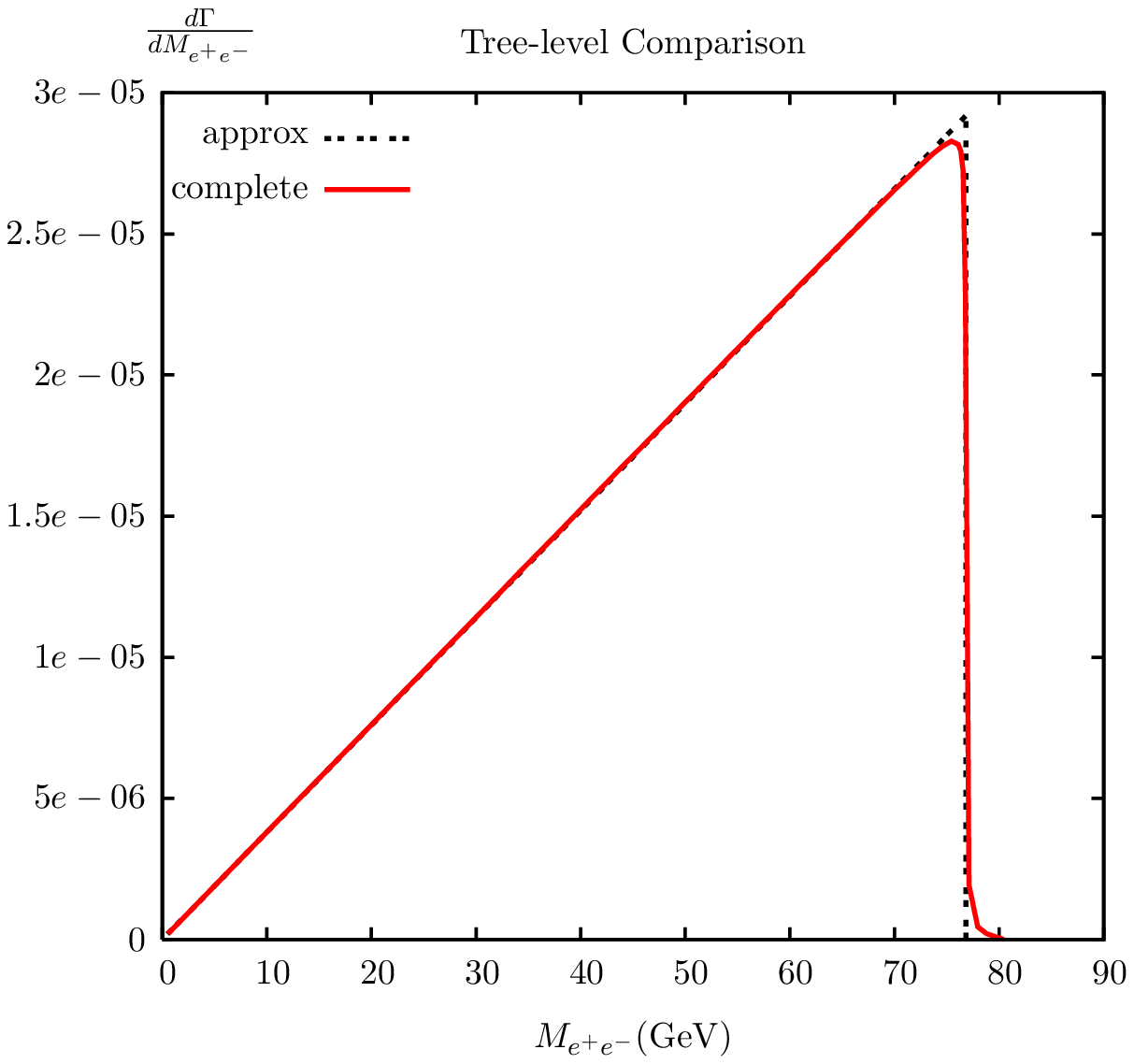}
\includegraphics[width=0.48\linewidth]{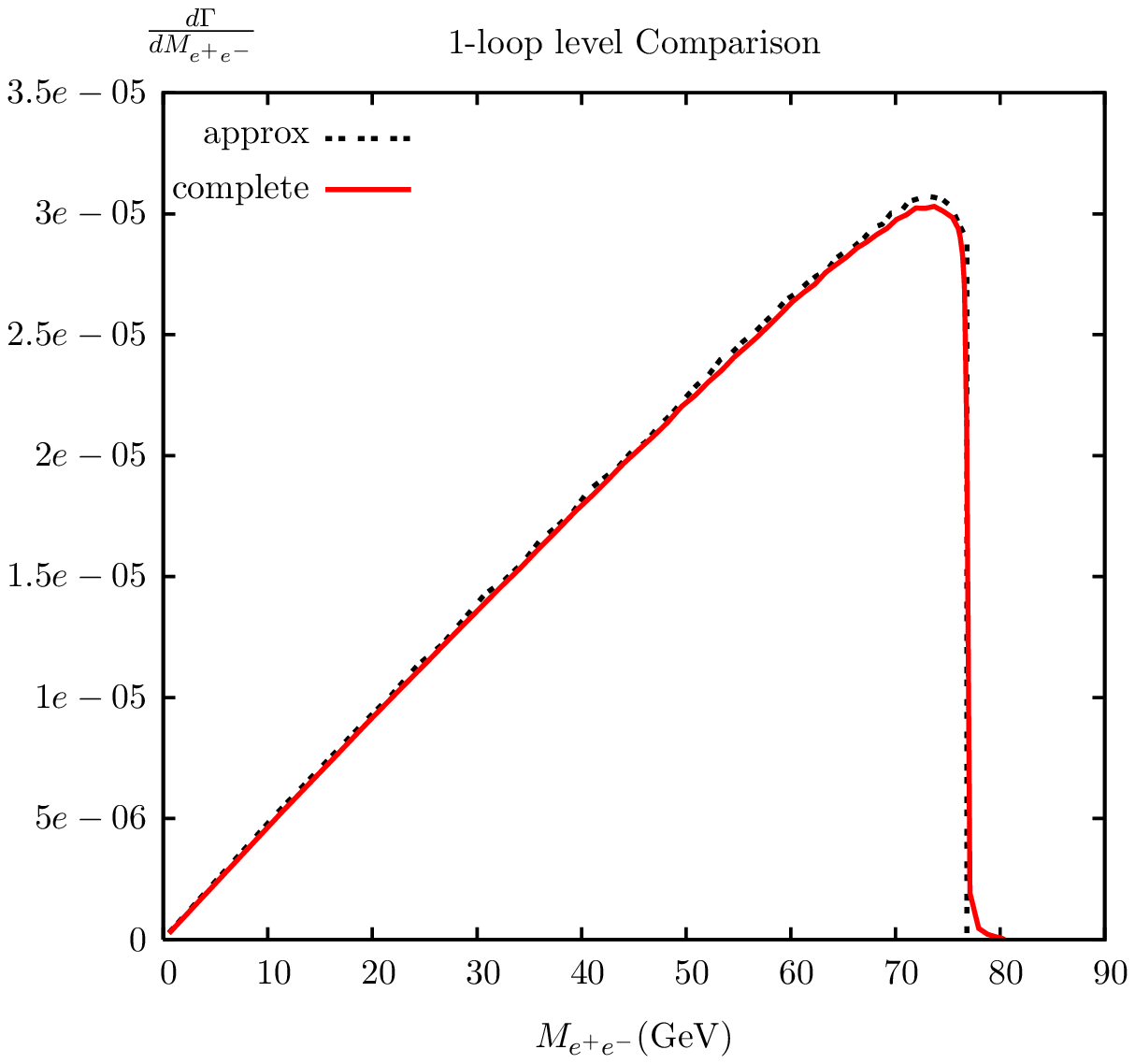}
\end{tabular}
\caption{The dilepton invariant mass $M_{e^+ e^-}$ distribution for the SPS1a
  parameter set (comparison of the complete calculation with the approximate
  calculation). \label{comparisonEE}}
\end{figure}
Fig.~\ref{SPS1aMee}, which shows a blow-up of the endpoint region, shows that
the peak of this distribution is then moved about $4$ GeV below the endpoint 
(\ref{Mee_end}) once higher-order corrections are included.
This is almost entirely due to contributions where a hard photon is emitted,
which takes away energy from the $e^+e^-$ system. This change of the shape of
the invariant mass distribution near the endpoint is important, since in
(simulated) experiments one needs a fitting function describing this
distribution in order to determine the location of the endpoint \cite{meefit}.
In Fig.~\ref{SPS1aMee} we used $\Delta \theta = 1^\circ$ in the definition of
collinear photons. In a real experiment, even photons emitted at somewhat
larger angles might be counted as contributing to the energy of the emitting
electron. In this case the change of the shape of the $M_{e^+e^-}$ 
distribution will be somewhat smaller.

In Fig.~\ref{comparisonEE} we compare the numerical results of the complete
calculation and the single-pole approximation at tree (left) and one-loop
level (right). At tree level the $M_{e^+e^-}$ distribution computed in the
single-pole approximation has an exactly triangular shape, with a sharp edge
at the endpoint (\ref{Mee_end}). This edge is smeared out a bit in the
complete tree-level calculation, which includes the full set of diagrams
shown in Fig.~1. As noted above, this edge is also softened considerably once
hard photon emission is included. The single-pole approximation therefore
works even better in the one-loop calculation. However, this excellent
agreement even for the differential decay width is partially accidental. The
agreement would become somewhat worse if the endpoints in two- and three-body
kinematics were further apart; this would happen if the mass of $\tilde l_1$
was close to the mass of either $\tilde \chi_2^0$ or to that of $\tilde
\chi_1^0$, since then one of the two square roots in (\ref{Mee_end}) would
become small.

\begin{figure}[t!]
\psfrag{u}{{\tiny $\mu$}}
\begin{tabular}{cc}
\includegraphics[width=0.48\linewidth]{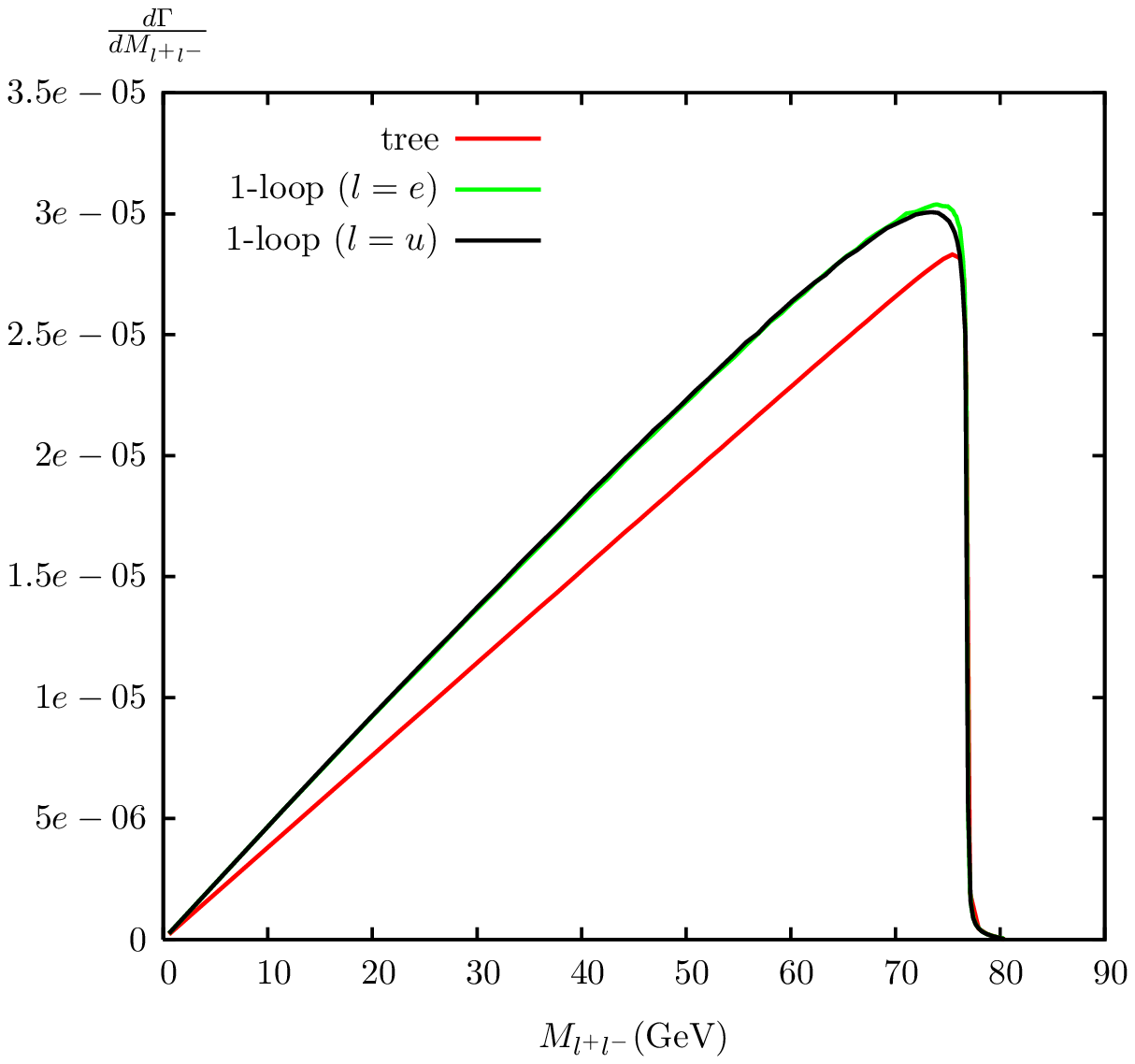}&
\includegraphics[width=0.46\linewidth]{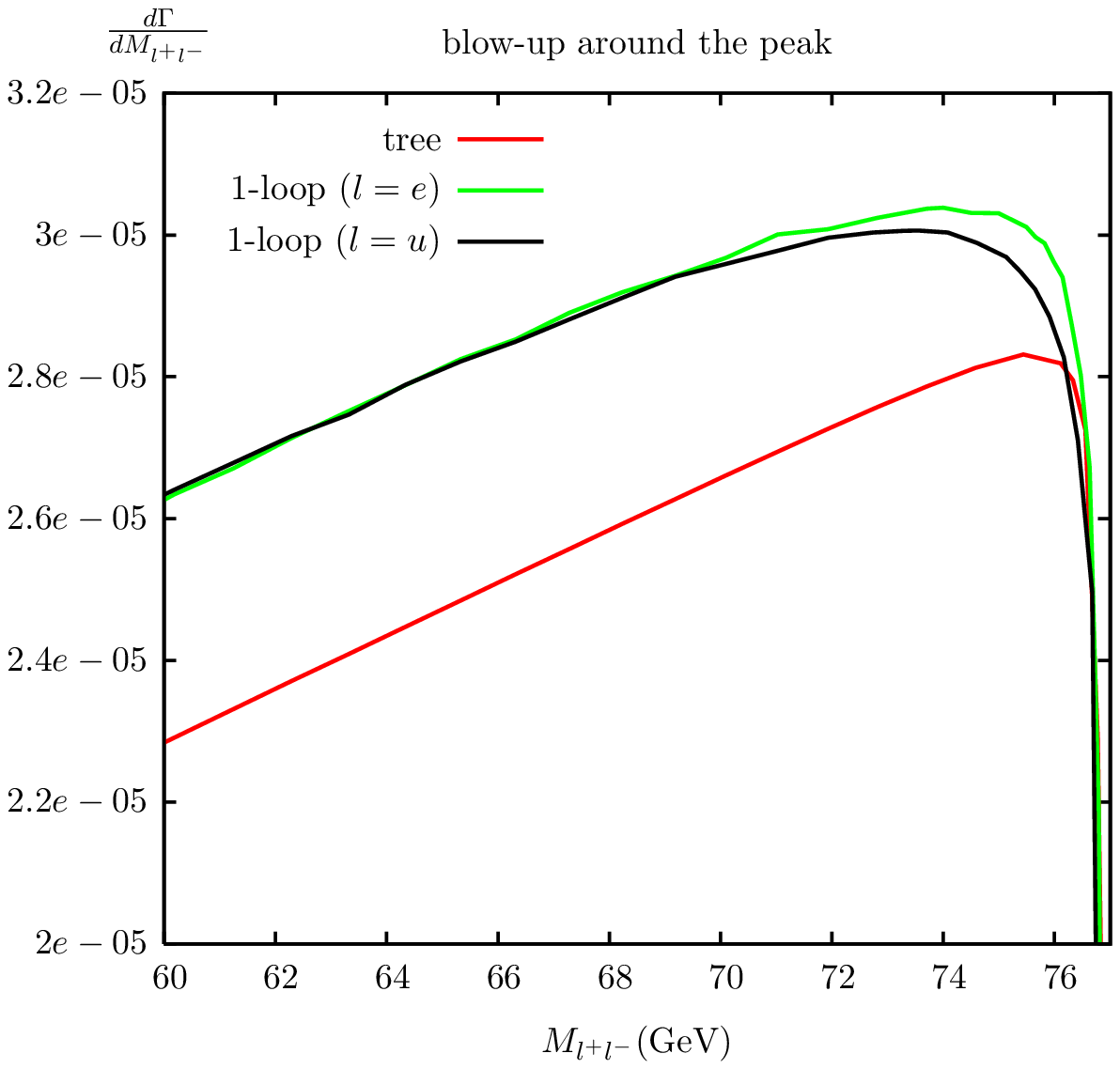}\\
\includegraphics[width=0.48\linewidth]{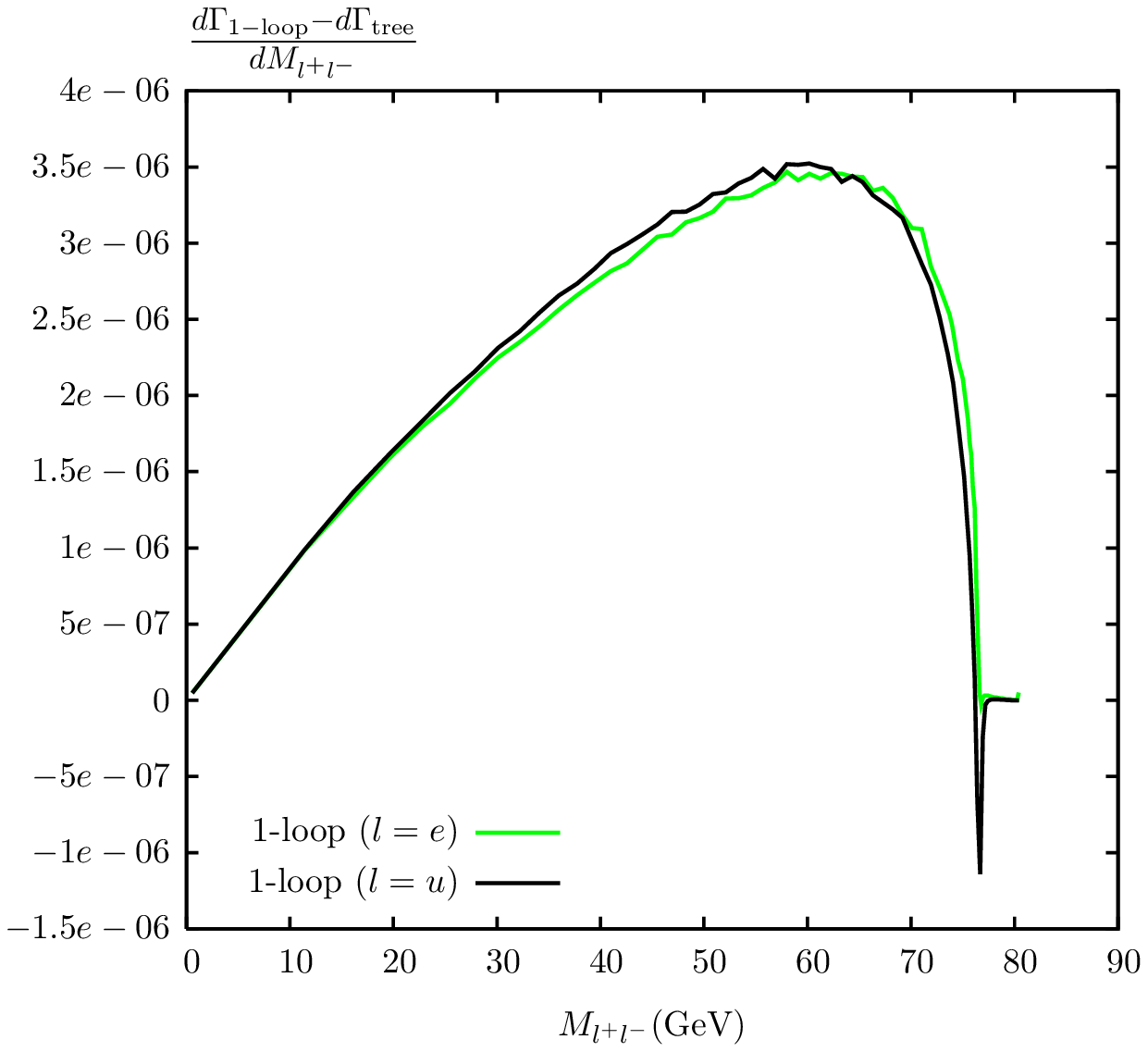}&
\includegraphics[width=0.46\linewidth]{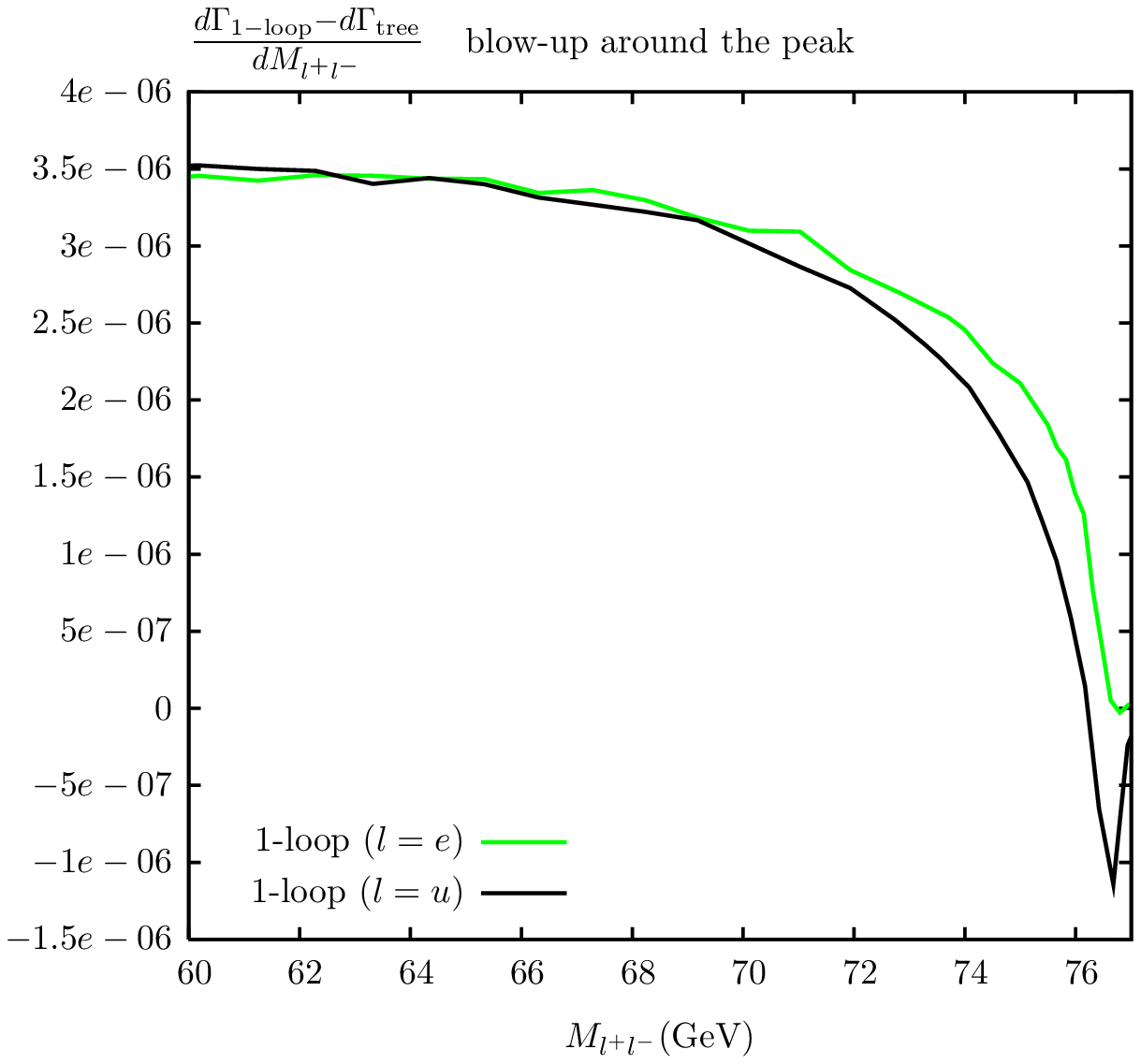}
\end{tabular}
\caption{The comparison of the dilepton invariant mass $M_{\mu^+ \mu^-}$ and 
$M_{e^+e^-}$ distribution for the SPS1a
  parameter set. \label{SPS1aMuu}}
\end{figure}

The comparison of the dilepton invariant mass $M_{\mu^+\mu^-}$ and $M_{e^+e^-}$
distributions is shown in Fig. \ref{SPS1aMuu}. In the upper frames we show 
the dilepton invariant mass $M_{l^+l^-} (l = e, \mu)$ distribution both at 
tree and one-loop level. Since the selectrons and smuons have 
equal masses and the light lepton mass $m_l~(l = e, \mu)$ is neglected 
except when it appears in the one-loop integrals,
their distributions are identical at tree level and different at one-loop level
due to the different treatment of the collinear-photon radiation. From these 
figures one obtains that at one-loop level the mass effect is larger 
near the endpoint than in other regions and the peak of the $M_{\mu^+\mu^-}$ 
distribution is shifted to lower invariant-mass values in comparison with the 
$M_{e^+e^-}$ distribution. We also show the relative one-loop corrections in 
the lower frames in Fig. \ref{SPS1aMuu}. The relative one-loop corrections 
from the $\mu^+\mu^-$ final state is smaller than that of the $e^+e^-$ final 
state in the upper invariant-mass region, while it is larger in the lower 
invariant-mass region. The main reason is that we add the momenta of collinear
photons to that of emitting electrons, but we do not do this for the 
collinear-photon radiation from muons. 
Hence the invariant mass $M_{\mu^+\mu^-}$ is reduced in comparison with 
$M_{e^+e^-}$. This leads to the shifting of events 
from the upper invariant-mass region to the lower invariant-mass region.  

\begin{figure}[htb]
\begin{tabular}{ll}
\includegraphics[width=0.48\linewidth]{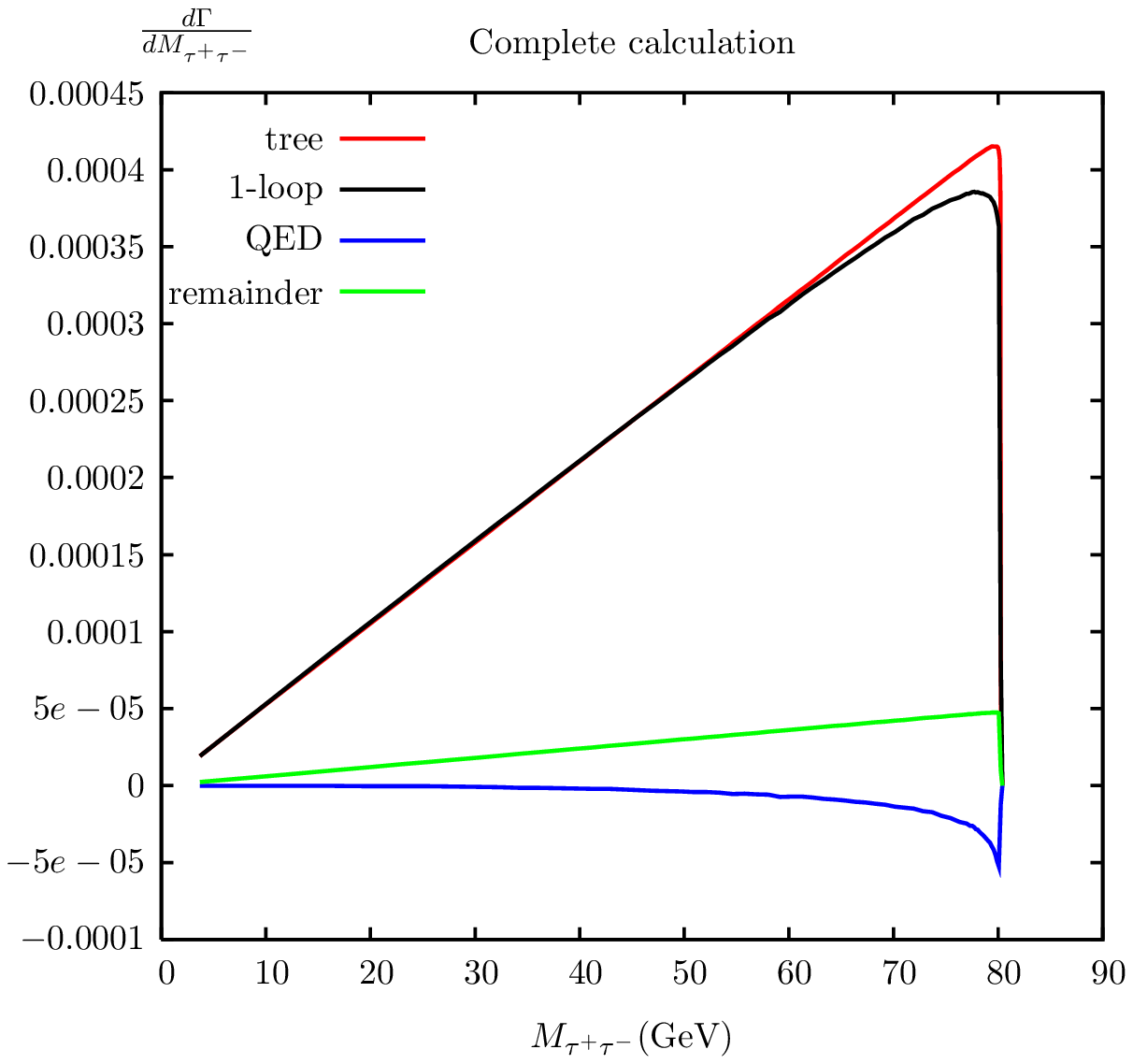}&
\includegraphics[width=0.45\linewidth]{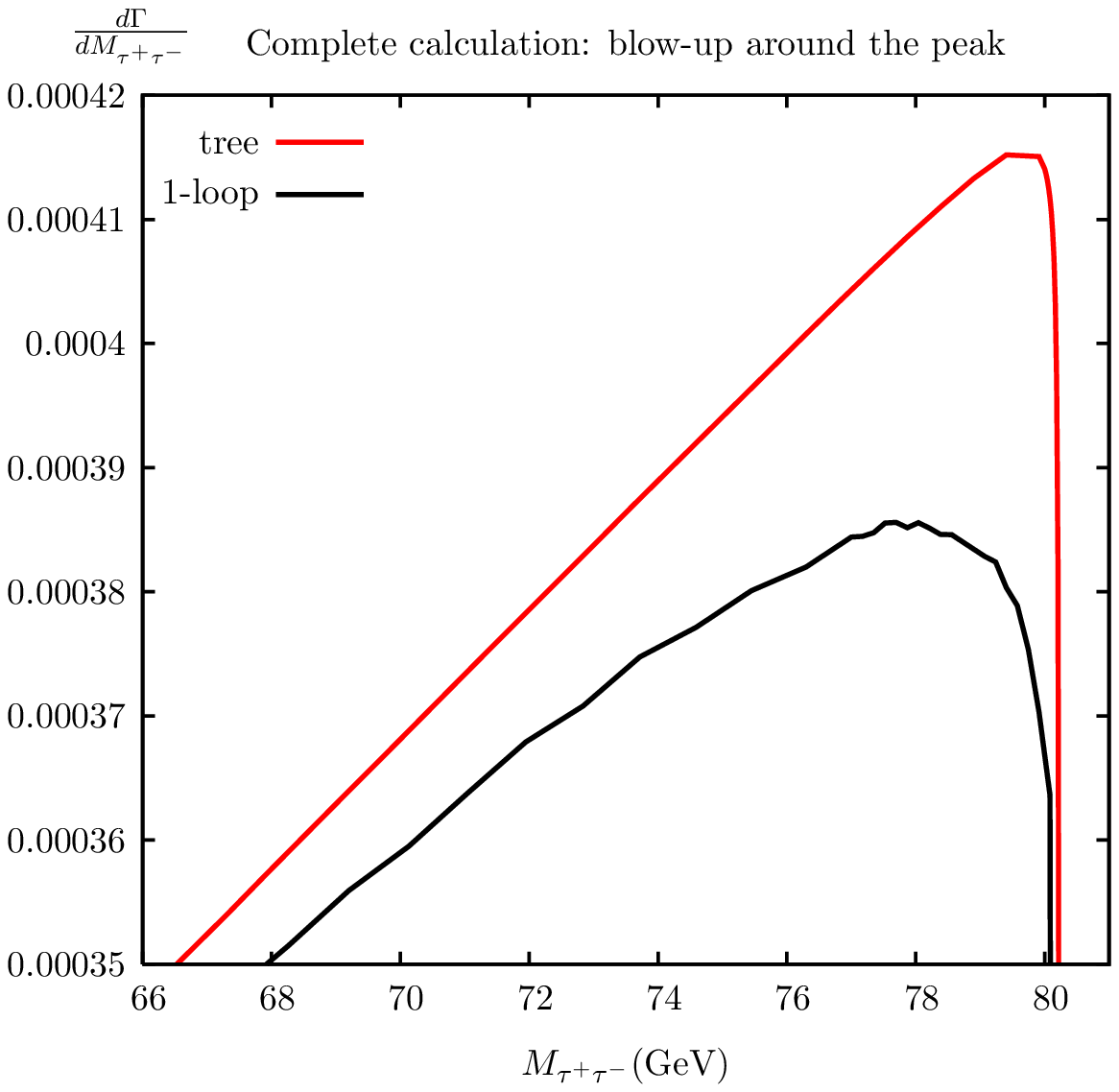}
\end{tabular}
\caption{The dilepton invariant mass $M_{\tau^+ \tau^-}$ distribution for the
  SPS1a parameter set (complete calculation). \label{SPS1aMll}}
\end{figure}

\begin{figure}
\begin{tabular}{cc}
\includegraphics[width=0.472\linewidth]{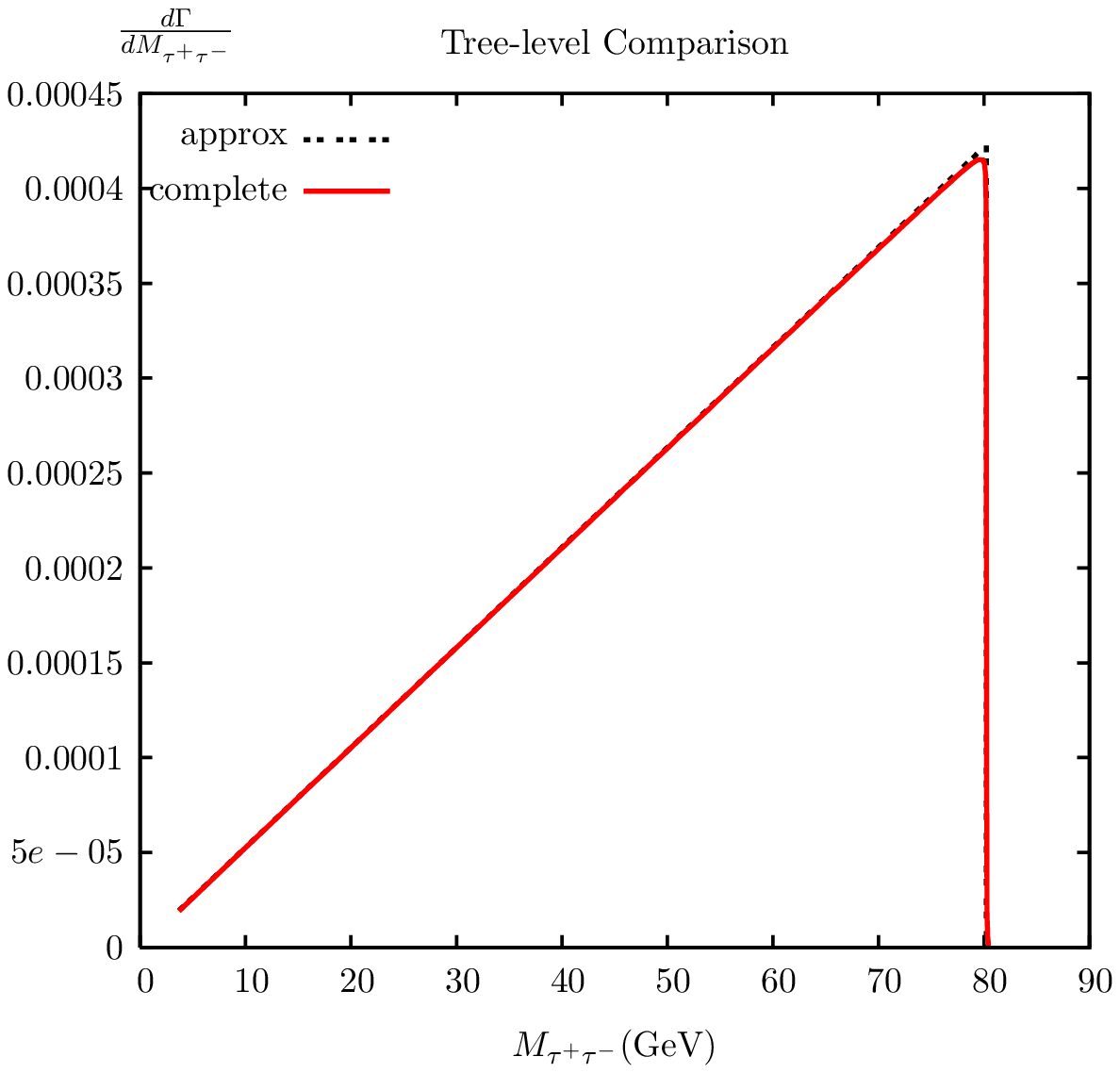}&
\includegraphics[width=0.472\linewidth]{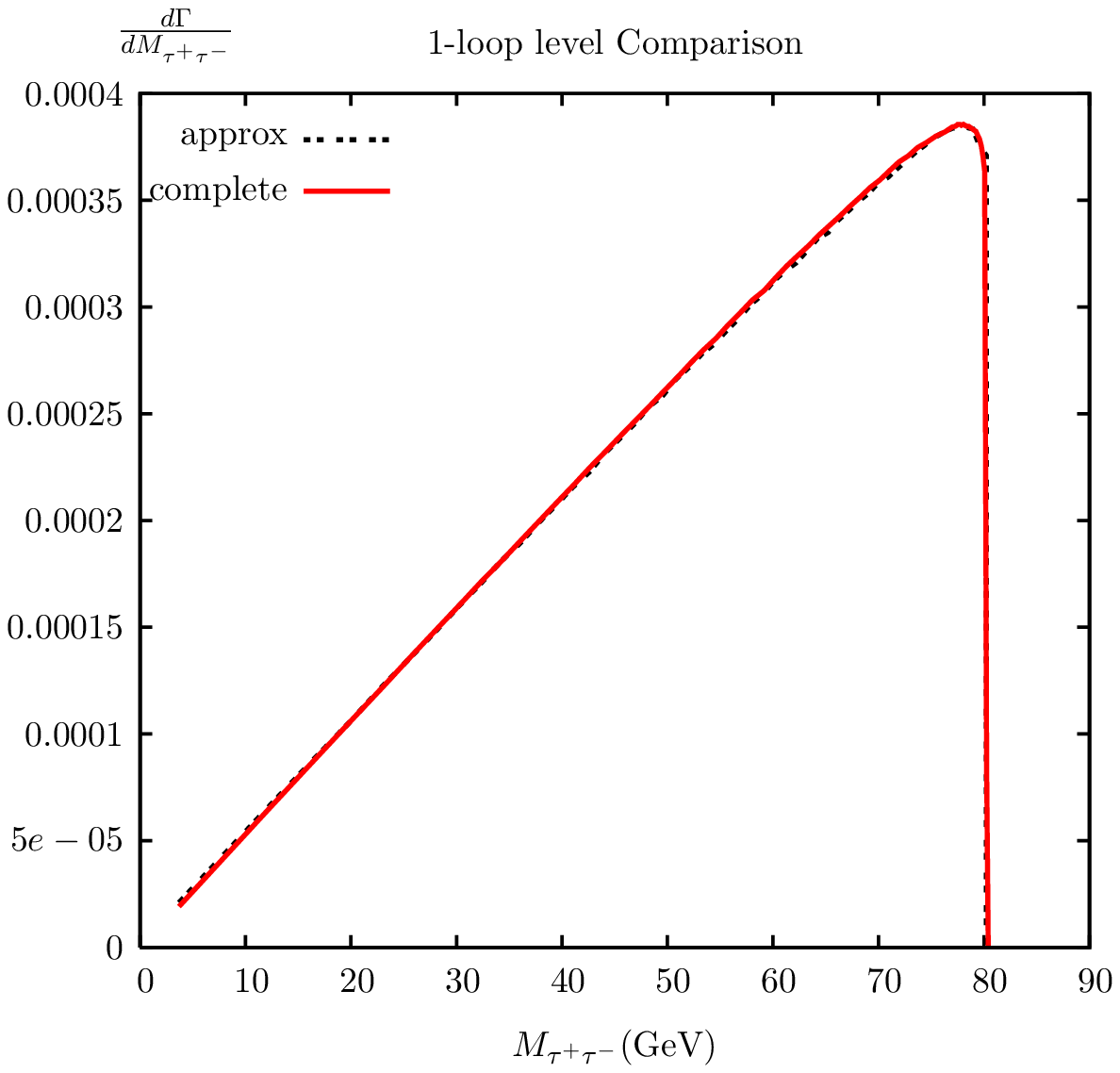}
\end{tabular}
\caption{The dilepton invariant mass $M_{\tau^+ \tau^-}$ distribution for the
  SPS1a parameter set (comparison of the complete calculation with the
  approximate calculation). \label{comparisonLL}} 
\end{figure}

The corresponding results for the $\tau^+ \tau^- \tilde \chi_1^0$ final state
are shown in Figs.~\ref{SPS1aMll} and \ref{comparisonLL}. Fig.~\ref{SPS1aMll}
shows that both the QED and, in particular, the non-QED corrections are
smaller in magnitude than for light leptons. In case of the QED contribution
this is essentially a mass effect. Our angular cutoff $\Delta \theta$ defining
the collinear region in (\ref{brems2}) is so small that even non-collinear
radiation off electrons or muons is still more likely than any hard radiation
off $\tau$ leptons; recall that we do not split hard radiation into collinear
and non-collinear contributions for $\tau^+ \tau^- \tilde \chi_1^0$ final
states.

The reduction of the non-QED corrections is even more dramatic. They amount
to about $+20\%$ for electrons and muons, but only to about $+6\%$ for tau
leptons. This difference stems from the fact that $\tilde l_1$ is a pure
$SU(2)$ singlet for $l=e,\, \mu$, since we neglect terms $\propto m_l$ in the
mass matrices of these sleptons. In contrast, $\tilde \tau_L - \tilde \tau_R$
mixing is quite significant, leading to a sizable $SU(2)$ doublet component of
$\tilde \tau_1$. Therefore $\tilde \chi_2^0$ decays into (real or virtual)
$\tilde l_1$ can only proceed through its small $U(1)_Y$ gaugino (bino)
component for $l=e, \, \mu$, while the large $SU(2)$ gaugino (neutral wino)
component also contributes for $l=\tau$. 
Moreover, the $\tilde\chi_1^\pm \tilde l_1^\mp \nu_l$ coupling,
which is involved in the virtual corrections, only exists for $l=\tau$.
The reason is that $\tilde l_1$ is a pure $SU(2)$ singlet for $l=e,\, \mu$ 
while $\tilde \tau_1$ has a sizeable $SU(2)$ doublet component, as explained 
above. In the limit of exact SUSY,
these contributions involve the $U(1)_Y$ and $SU(2)$ gauge couplings,
respectively, which renormalize (and run) quite differently. If there are
significant differences between masses of supersymmetric particles, 
the differences between the true gauge couplings and these 
gaugino-lepton-slepton couplings also becomes significant \cite{inocoup};
note that in the SPS1a scenario, squarks (which contribute to various 
two-point functions) are about three times heavier than $\tilde \chi_2^0$. 
Finally, the $\tilde \chi_2^0 \tilde \tau_1
\tau$ vertex also receives non-negligible contributions which, again in the
limit of exact SUSY, are proportional to the $\tau$ Yukawa coupling
\cite{MSSM}. As a result of the reduced non-QED corrections, the total
correction is now {\em negative}, especially for large values of $M_{\tau^+
  \tau^-}$.

The endpoint region of the $M_{\tau^+ \tau^-}$ distribution from the complete
calculation is shown in the right panel of Fig.~\ref{SPS1aMll}. We see that
the peak of the distribution is shifted downwards by about $2$ GeV once
higher-order corrections are included. A shift of this magnitude may be
significant, even though the $\tau^+ \tau^-$ invariant mass is in general
difficult to measure accurately, due to the presence of $\nu_\tau$
(anti-)neutrinos in the $\tau$ decay products, which escape detection.

In Fig.~\ref{comparisonLL} predictions from the complete calculation are
compared to those from the single-pole approximation. In this case we find
almost perfect agreement even in the endpoint region, both at tree level and
after including one-loop corrections. The reason is that for SPS1a,
$m_{\tilde \tau_1}$ happens to be very close to $\sqrt{ m_{\tilde \chi_1^0}
m_{\tilde \chi_2^0} }$. Performing the replacement $m_{\tilde e_1} \rightarrow
m_{\tilde \tau_1}$ in (\ref{Mee_end}) shows that the endpoints of the $\tau^+
\tau^-$ distributions in two- and three-body kinematics practically coincide.

\begin{table}[b!]
\begin{center}
\begin{tabular}{|l|l|l|}\hline
decay mode & tree-level width(MeV), \ \ Br&  one loop-level width(MeV), Br\\
\hline 
$e^{-} e^{+}\tilde{\chi}_1^0$ & $1.123$ (1.122), \hspace*{2.cm}$5.9\%$ &
$1.297$ (1.294), \hspace*{2.cm}$6.7\%$  \\ \hline 
$\mu^{-} \mu^{+}\tilde{\chi}_1^0$ & $1.123$ (1.122), \hspace*{2.cm}$5.9\%$ &
$1.297$ (1.294), \hspace*{2.cm}$6.7\%$\\ \hline 
$\tau^{-} \tau^{+}\tilde{\chi}_1^0$ & 16.870 (16.933), \hspace*{1.6cm}$88.0\%$
& 16.595 (16.646), \hspace*{1.6cm}$86.2\%$\\ \hline 
$\nu_e \bar \nu_e\tilde{\chi}_1^0$ & $0.012$ \hspace*{2.8cm}$$ & $0.012$
\hspace*{1.8cm}$$\\ \hline 
$\nu_\mu \bar \nu_\mu \tilde{\chi}_1^0$  & $0.012$ \hspace*{2.8cm}$$ & 0.012
\hspace*{1.8cm}$$  \\ \hline 
$\nu_\tau \bar \nu_\tau\tilde{\chi}_1^0$ & 0.013 \hspace*{2.8cm}$$ & $0.013$
\hspace*{1.8cm}$$\\ \hline 
$q \bar q\tilde{\chi}_1^0~(q\neq t)$& 0.015 \hspace*{2.8cm}$$ & $0.015$
\hspace*{1.8cm}$$\\ \hline
total width & 19.168 & 19.241\\ \hline
\end{tabular}
\end{center}
\caption{Partial widths of different $\tilde{\chi}_2^0$ decay modes and
  the branching ratios of its visible decays for the SPS1a parameter set. The
  numbers in parentheses give the corresponding partial widths calculated in
  the single-pole approximation.
\label{twobodytotal}}
\end{table}
The partial widths of the different $\tilde{\chi}_2^0$ decay modes and the 
branching ratios of its visible leptonic decays are listed in Table
\ref{twobodytotal}, where the numbers in the parentheses are obtained from the 
approximate calculations. We find $\sum_l \Gamma(\tilde \chi_2^0 \rightarrow 
\tilde \chi_1^0 \nu_l \bar\nu_l) \ll \sum_l \Gamma(\tilde \chi_2^0 \rightarrow
\tilde \chi_1^0 l^+ l^-)$. This is not surprising, since the charged lepton
final state is accessible via on-shell $\tilde l_1$ intermediate state,
whereas for the neutrino final state all exchanged particles are
off shell. Since squark masses are near 500 GeV in SPS1a scenario, hadronic
final states contribute even less than neutrinos do.

\begin{figure}[t!]
\psfrag{u}{{\tiny $\mu$}}
\begin{center}
\includegraphics[width=0.6\linewidth]{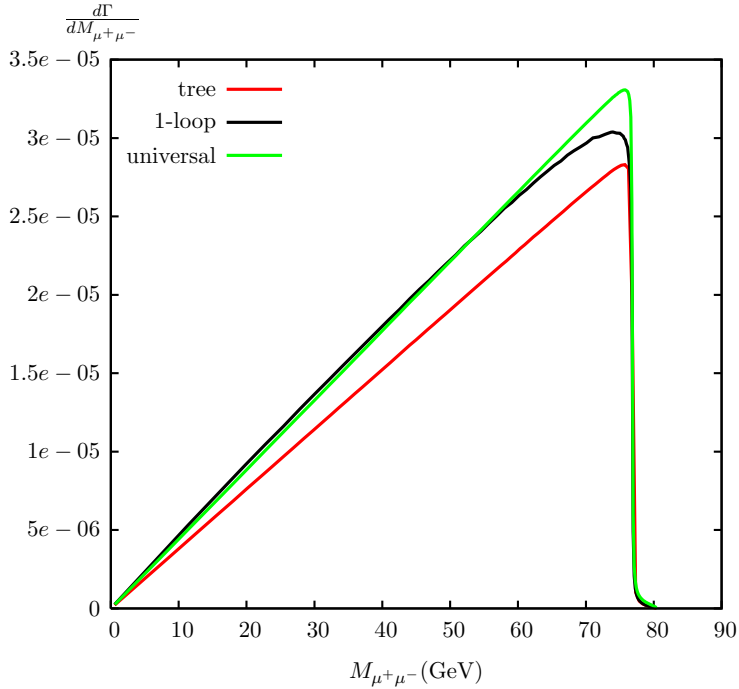}
\end{center}
\caption{Dilepton invariant mass $M_{\mu^+ \mu^-}$ 
from an approximate calculation with only universal 1-loop contributions
via effective couplings, in comparison with the tree-level 
and the complete 1-loop results.
SPS1a parameter set. }
\label{universal}
\end{figure}

From the results in Table~\ref{twobodytotal} one concludes:
\begin{itemize}
  
\item The main decay mode of $\tilde{\chi}_2^0$ is $\tilde{\chi}_2^0
  \rightarrow \tau^{-} \tau^{+}\tilde{\chi}_1^0$. Its branching ratio is about
  $88.0\%$ at tree-level, $86.2\%$ at one-loop level. This mode dominates
  partly because of the lower mass of $\tilde \tau_1$ as compared to $\tilde
  e_1$ (133.0 GeV vs 142.7 GeV). Even more important is that the $\tilde
  \chi_2^0 \tilde \tau_1 \tau$ coupling is much stronger than the $\tilde
  \chi_2^0 \tilde e_1 e$ coupling, which in turn is due to significant $L-R$
  mixing, which only exists in the $\tilde \tau$ sector, as explained above.
  
\item The total $\tilde \chi_2^0$ decay width is enhanced by 0.4\% when 
  one-loop corrections are included. Such modest corrections are typical 
  in the absence
  of large enhancement factors (e.g., large logarithms). This overall
  perturbative stability confirms that our choice of renormalization scheme,
  and of the electroweak input parameters listed in Appendix A, is indeed
  rather well suited for the task at hand.\footnote{Of course, the total width
  after the inclusion of one-loop corrections is scheme independent, up to
  unknown two-loop correction terms. However, the relative size of the
  one-loop corrections does depend on the chosen scheme.}

\item One-loop corrections enhance the partial width and the branching ratio
  of $\tilde{\chi}_2^0 \rightarrow l^- l^+ \tilde{\chi}_1^0 \ (l = e, \mu)$
  decays by $15.5\%$ and $13.6\%$, respectively. This results from the large 
  size of the positive non-QED corrections depicted in Fig.~3. Much of these
  corrections can probably be absorbed into an appropriately defined running
  $\tilde \chi_2^0 \tilde l_1 l$ coupling. This is illustrated in 
  Fig.~\ref{universal}, which compares the $M_{l^+l^-}$ distribution computed 
  including only the ``universal corrections" defined in Ref.~\cite{guasch} 
 (see also~\cite{inocoup}) with the tree-level and full one-loop results. 
  We see that the residual non-universal corrections are relevant only close 
  to the edge of the lepton pair distribution, where real photon emission is 
  most important. Since this result is for a specific scenario, a more 
  comprehensive analysis might be appropriate.

\item The single-pole approximation reproduces the integrated partial widths
  to about 0.3\% accuracy. This agreement is even better than in the 
  $M_{l^+l^-}$ distribution shown in Figs.~4 and 6. In fact, 
  from (\ref{propid}) and the discussion at the end of Sec.~4.1.2 
  one might expect better agreement for the integrated
  partial width than for (some) kinematical distributions.
\end{itemize}

\subsection{Numerical results for pure three-body decays}
We also investigated the effect of higher-order corrections on leptonic
$\tilde \chi_2^0$ decays for a scenario where $\tilde \chi_2^0$ does not have
any two-body decay modes. To that end we again use the SPS1a parameter set,
except that the soft SUSY-breaking parameters in the slepton mass matrix are 
set to
\begin{eqnarray} \label{newslep}
m_{\tilde{l}_L} = 230 \ {\rm GeV}\, , \ \ m_{\tilde{l}_R} = 183 \ {\rm GeV},
 \hspace*{3mm}l = e, \,
\mu,\, \tau\, . 
\end{eqnarray}
The masses of the relevant neutralinos and sleptons in this modified SPS1a 
parameter set are listed in Table~\ref{spstaboff} where one finds that 
$\tilde{\chi}_2^0$ has to undergo a pure three-body decay. Therefore we do not 
have to introduce complex slepton masses in the one-loop functions. 
Apart from this simplification, the calculation is very similar to the 
``complete'' calculation described in Sec.~4.1.
\begin{table}[htb] 
\begin{center}
\begin{tabular}{|l|l|l|l|l|l|l|l|}\hline
particle & \ $\tilde{\chi}_2^0$ & $\ \tilde{\chi}_1^0$ &
$\tilde{e}_1 \, (\tilde{\mu}_1)$ & $\tilde{e}_2 \, (\tilde{\mu}_2)$ &
\ \ $\tilde{\tau}_1$ & \ \ $\tilde{\tau}_2$ & $\tilde{\nu}_l
(l = e\, ,\mu\, ,\tau)$  \\ \hline
mass (GeV) & 176.6 & 96.2 & 187.9 &  234.9 & 182.3 & 239.2 & \ \ \ 221.0  \\
\hline 
\end{tabular}
\end{center}
\caption{Masses of the relevant neutralinos and sleptons for the modified 
SPS1a \mbox {parameter~set.}\label{spstaboff}}
\end{table}

\begin{figure}[b!]
\begin{tabular}{ll}
\includegraphics[width=.473\linewidth]{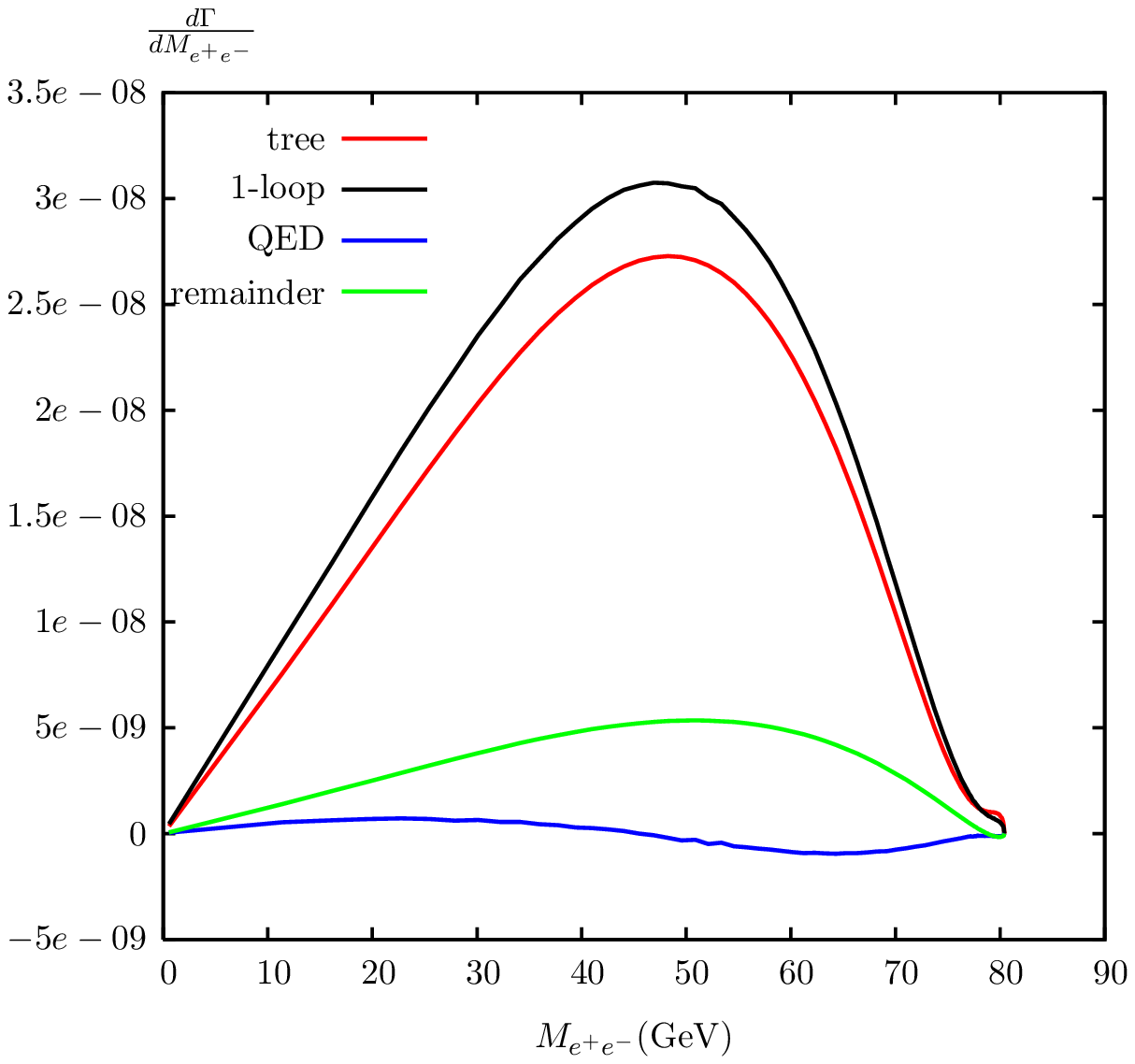} &
\includegraphics[width=.473\linewidth]{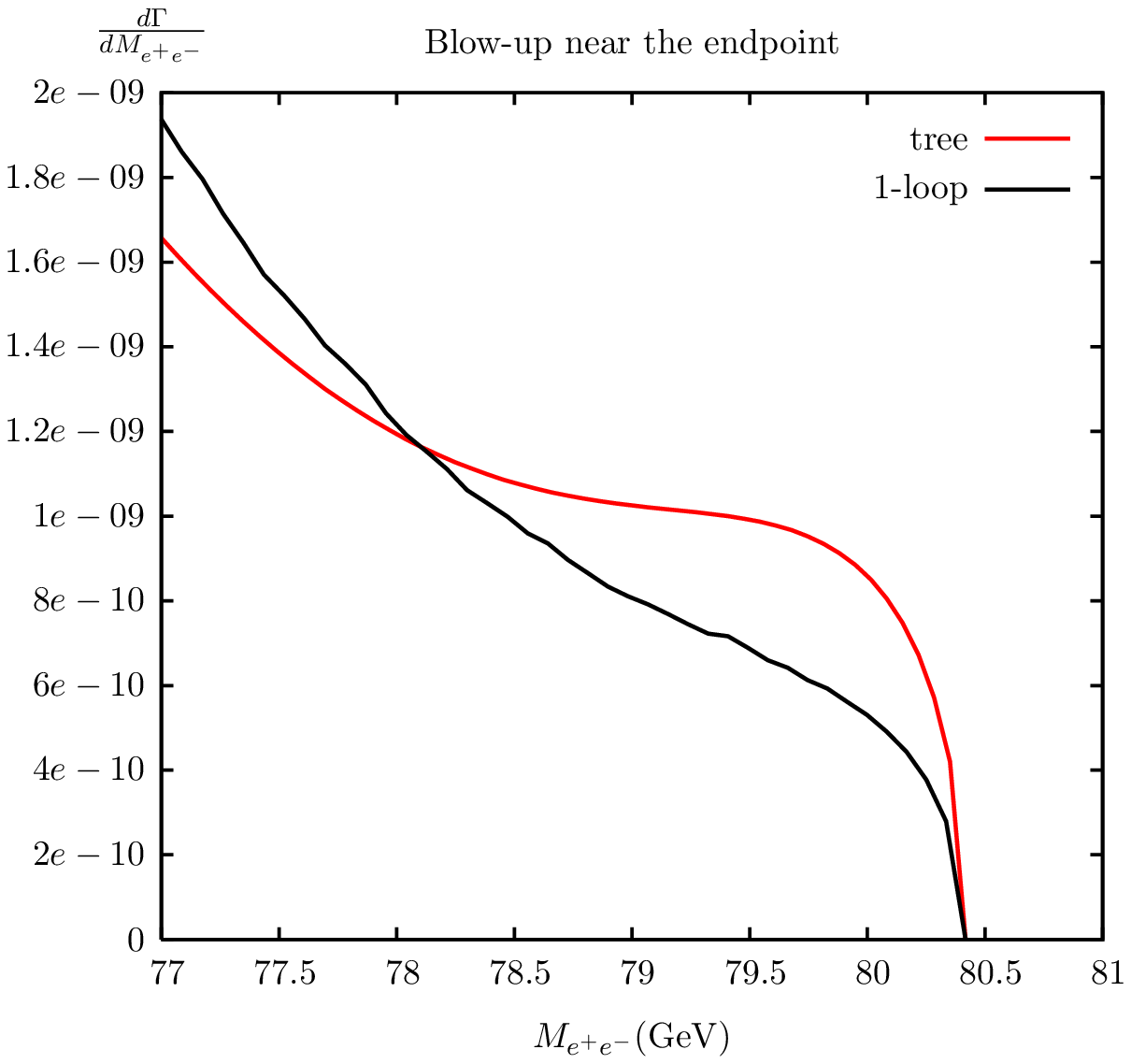}
\end{tabular}
\caption{The $\tilde \chi_2^0$ decay width differential in the dilepton
  invariant mass $M_{e^+ e^-}$ in the case of genuine three-body
  decay. \label{Mee}} 
\end{figure}

\begin{figure}[htb]
\begin{tabular}{ll}
\includegraphics[width=.49\linewidth]{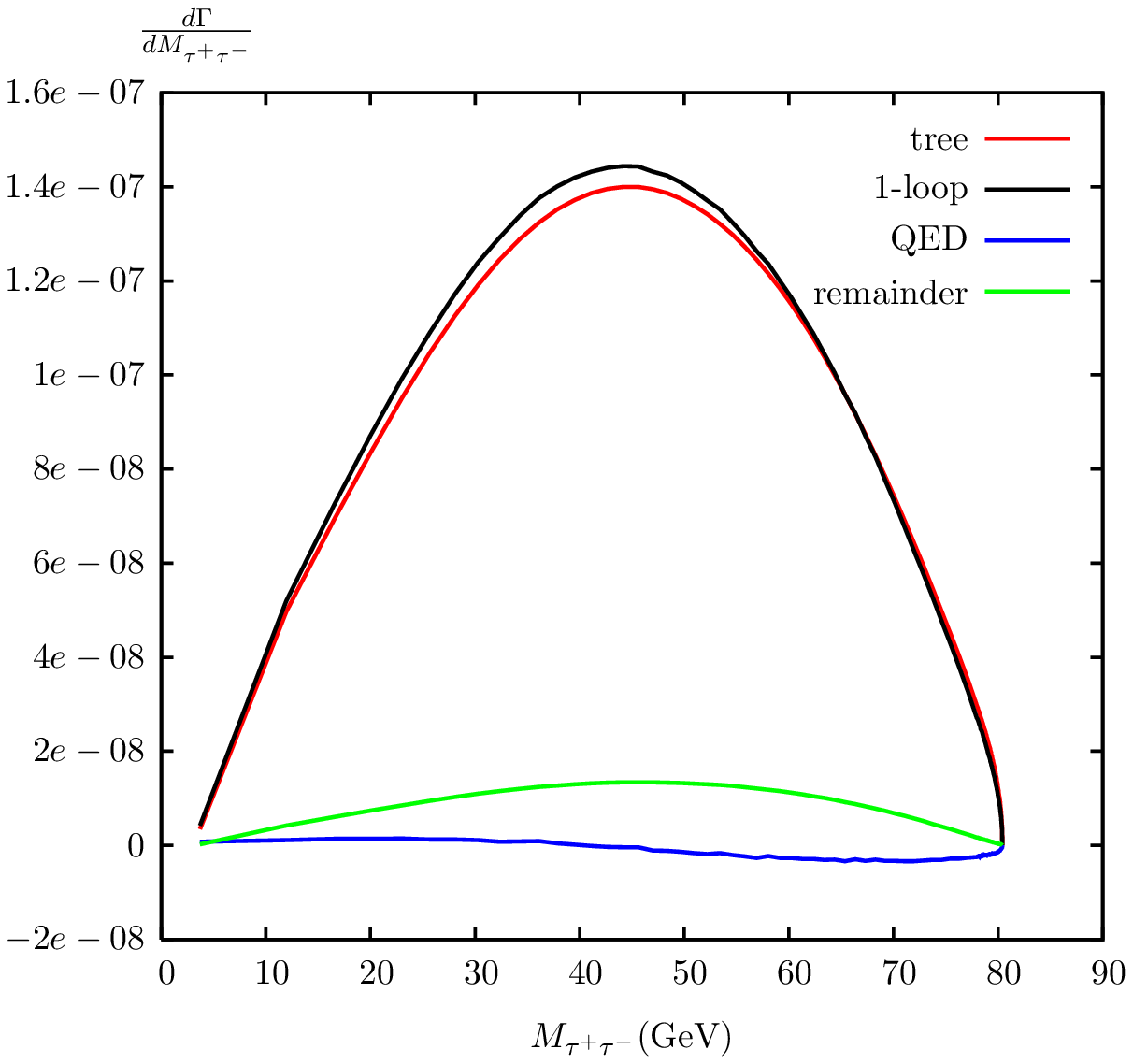} &
\includegraphics[width=.45\linewidth]{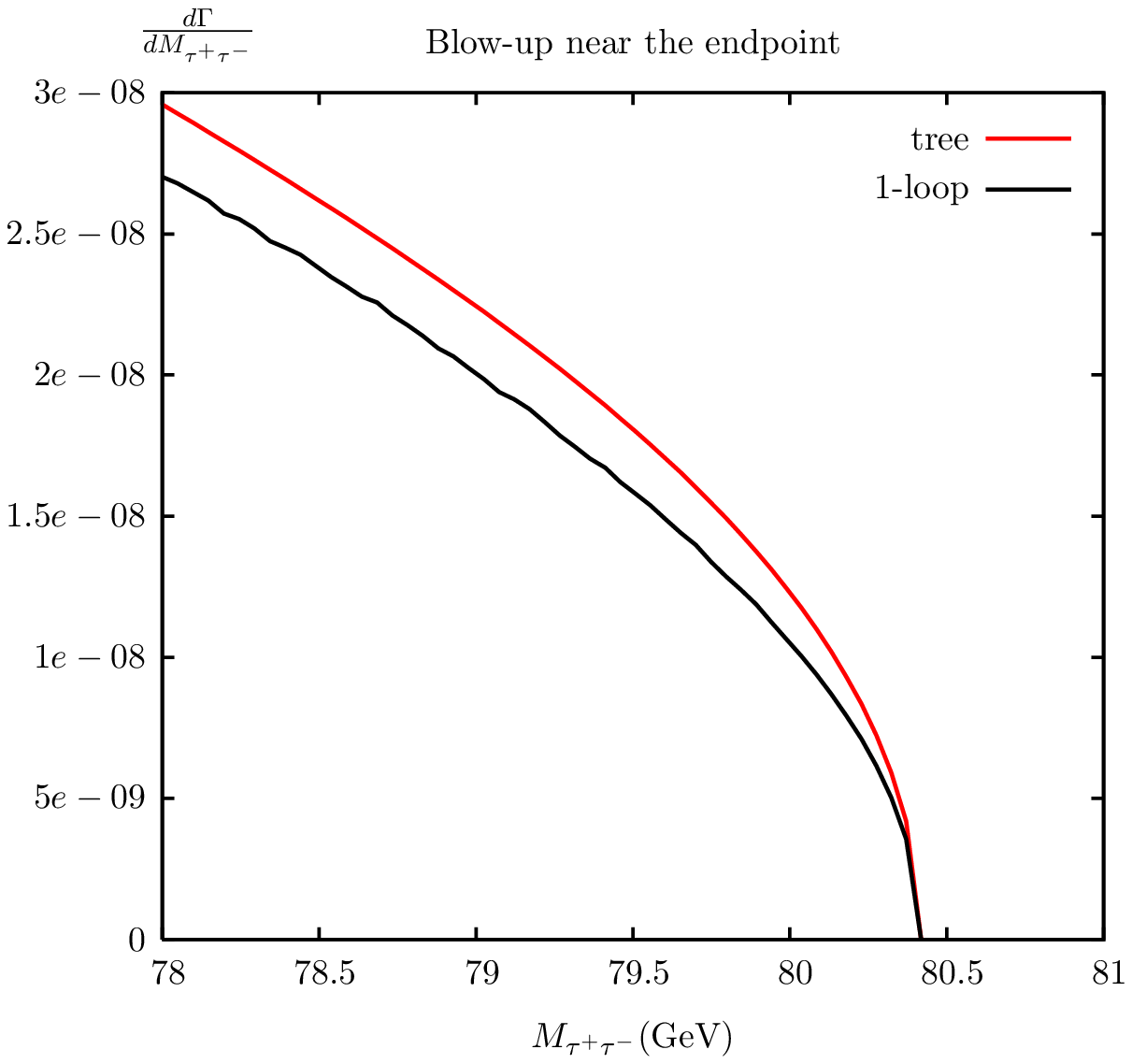}
\end{tabular}
\caption{The $\tilde \chi_2^0$ decay width differential in the di$-\tau$
  invariant mass $M_{\tau^+ \tau^-}$ in the case of genuine three-body
  decay. \label{Mll}}
\end{figure}

The dilepton invariant mass $M_{e^+ e^-}$ and $M_{\tau^+ \tau^-}$ distributions
are shown in Figs.~\ref{Mee} and \ref{Mll}, respectively. \
At tree level the $M_{e^+e^-}$ distribution shows
a small peak near its upper endpoint from the exchange of nearly on-shell $Z$
bosons. Since the QED and non-QED corrections are very small and negative in 
this region, this peak is
less pronounced once one-loop corrections are included. This is of some
significance, since the shape of this distribution can now be used to infer
the strengths of various contributing diagrams, which in turn provides
information on slepton masses and neutralino mixing \cite{noya,matchev}. Since
$\tilde \tau$ exchange is much enhanced relative to $\tilde e$ exchange, 
one cannot see any contributions of $Z$-exchange even at tree level
from the $M_{\tau^+ \tau^-}$ distribution.
Moreover we can observe that the invariant mass $M_{e^+ e^-}$ and 
$M_{\tau^+ \tau^-}$ distributions have a rather
sharp edge at their endpoints. These edges are again softened by real photon
emission, but remain quite distinct. This should facilitate the experimental
determination of the endpoint, and hence the measurement of $m_{\tilde
  \chi_2^0} - m_{\tilde \chi_1^0}$.

\begin{figure}[t!]
\psfrag{u}{{\tiny $\mu$}}
\begin{tabular}{cc}
\includegraphics[width=0.485\linewidth]{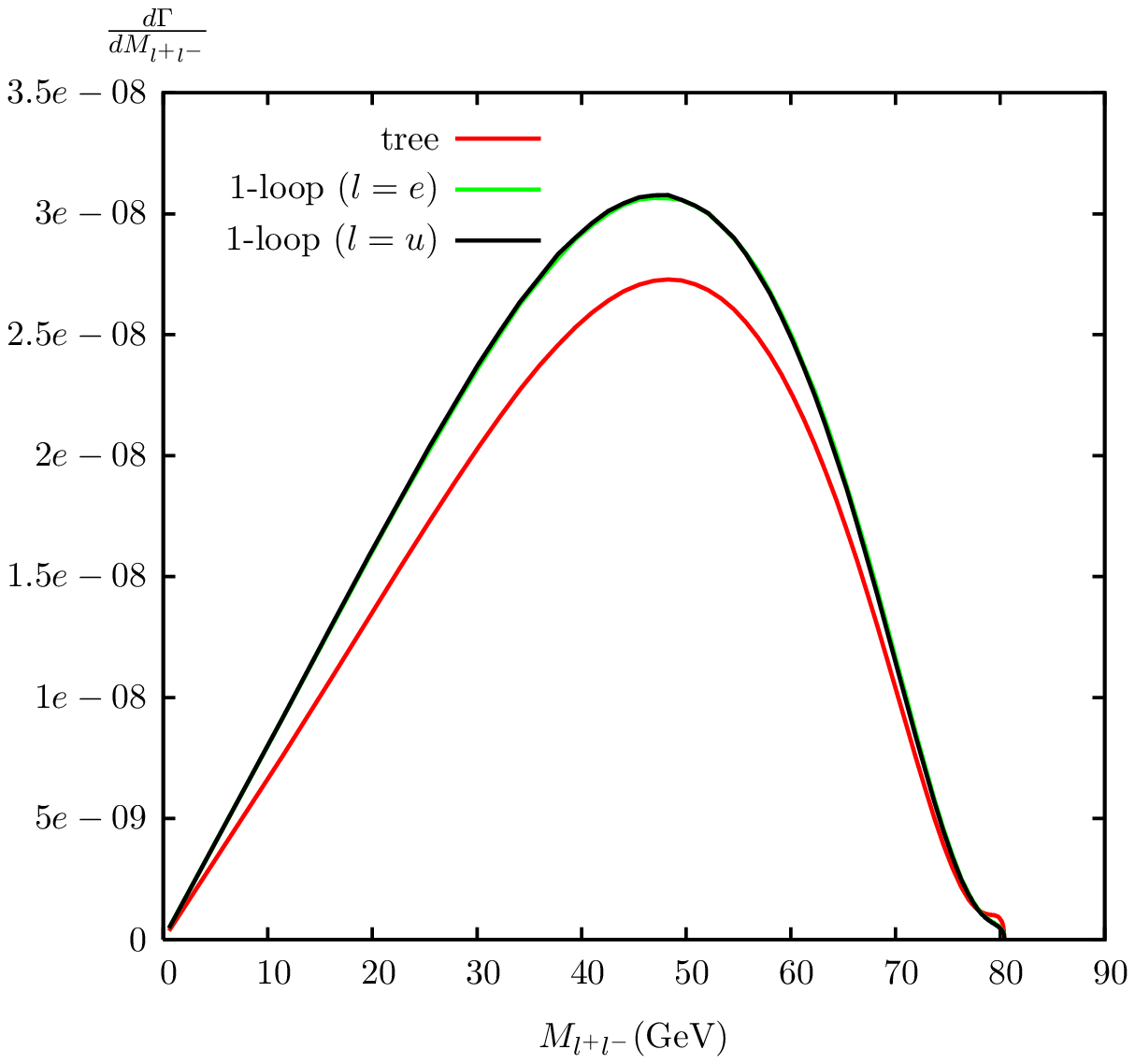}&
\includegraphics[width=0.46\linewidth]{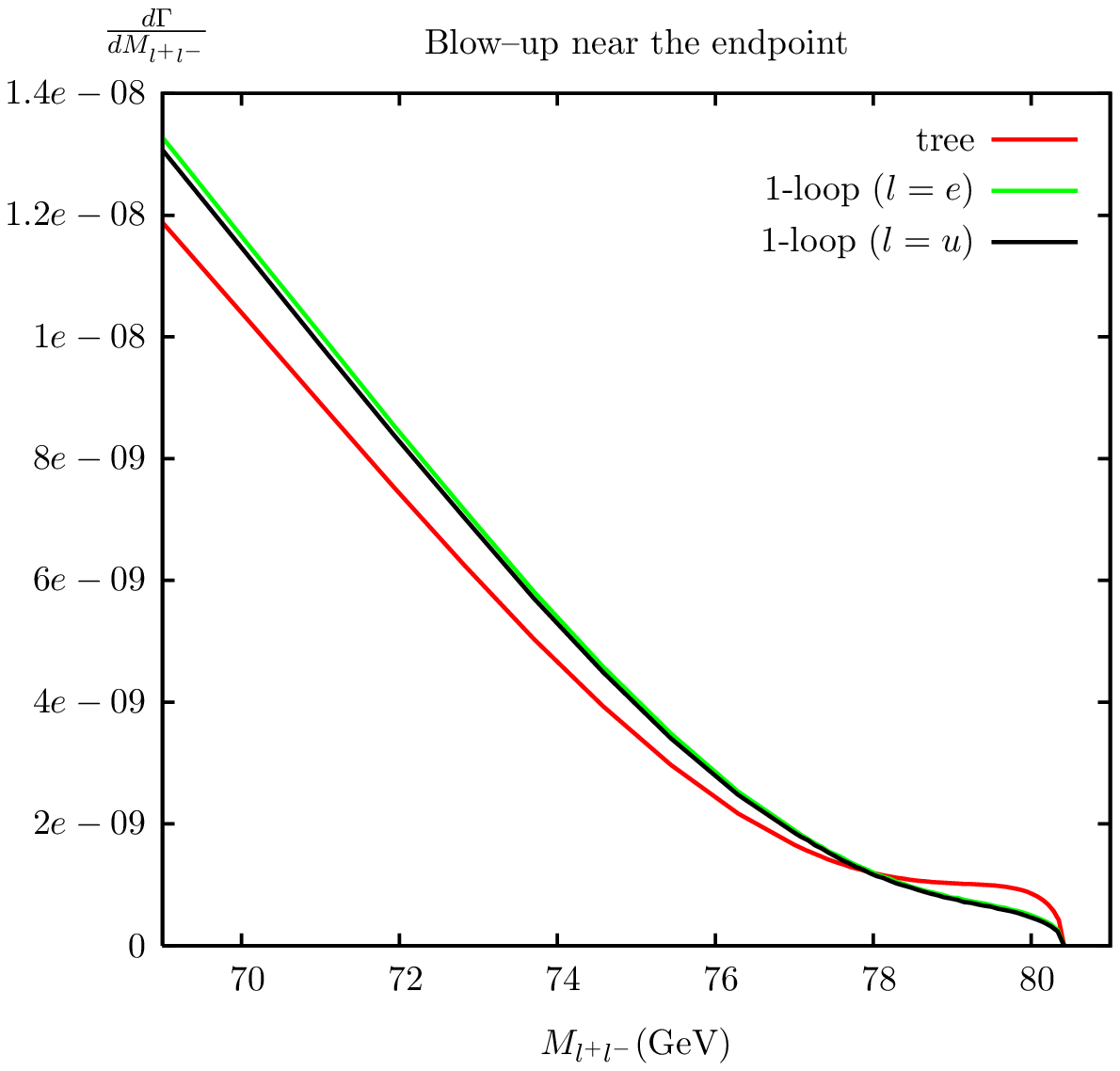}\\
\includegraphics[width=0.485\linewidth]{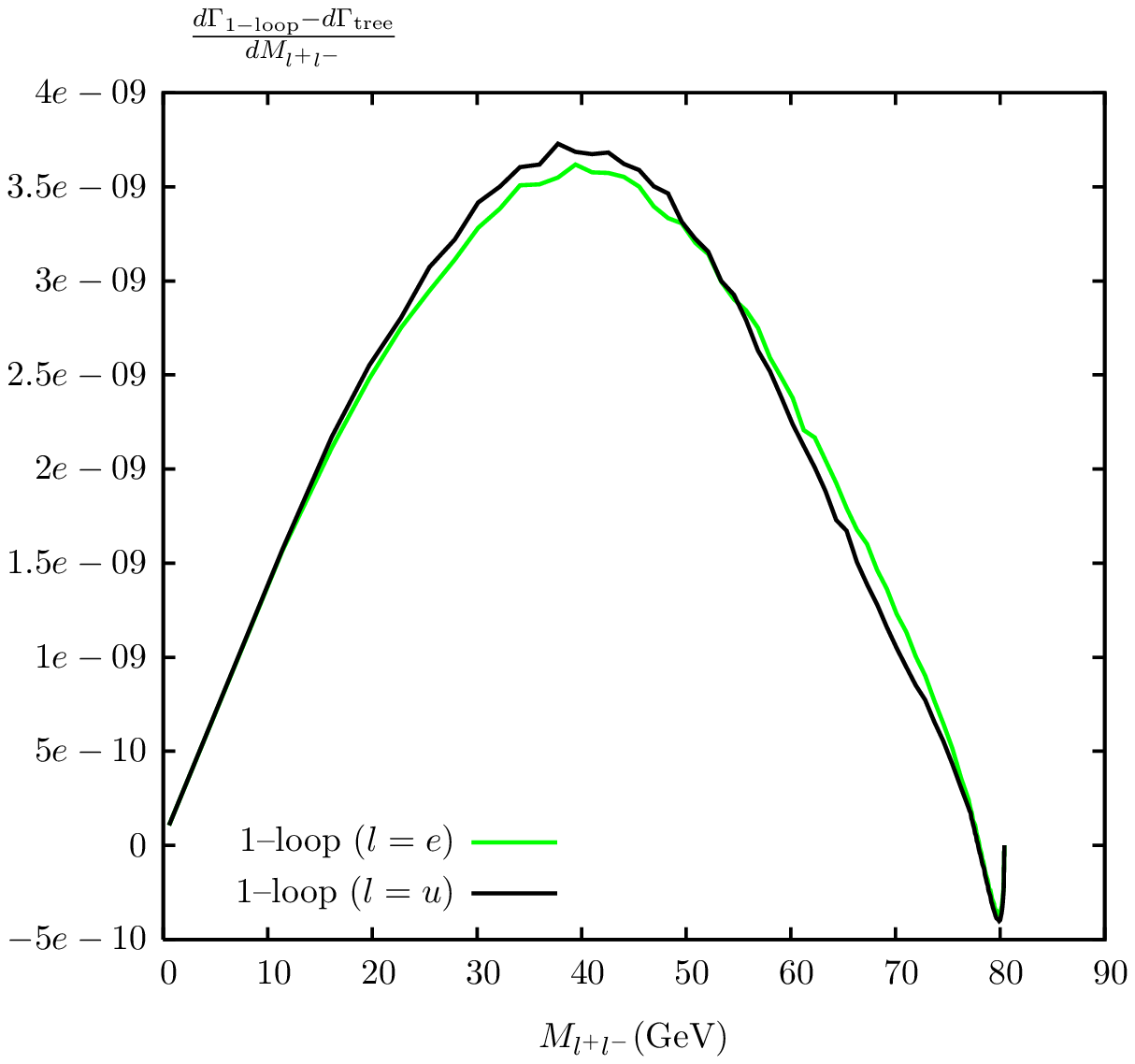}&
\includegraphics[width=0.46\linewidth]{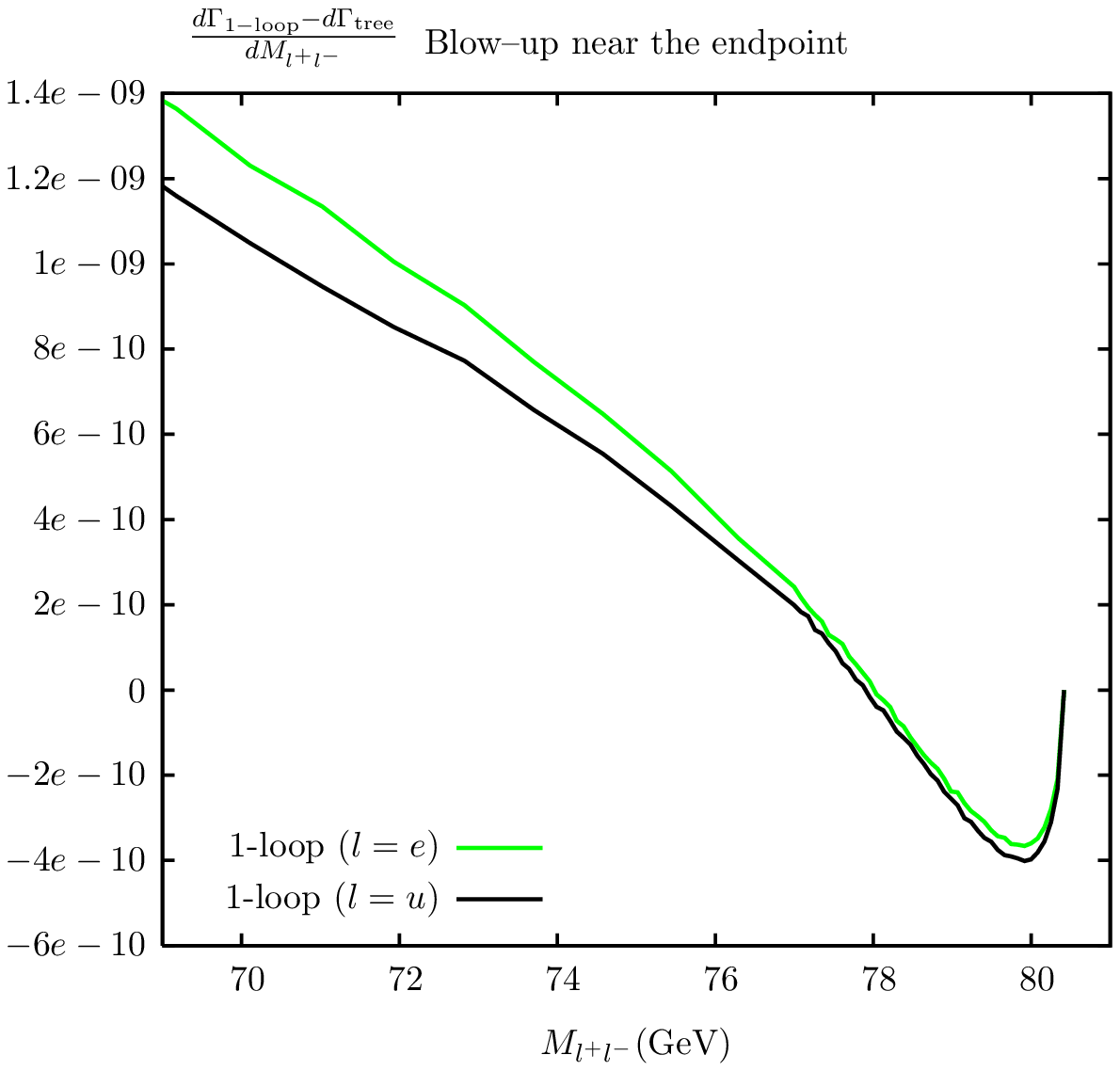}
\end{tabular}
\caption{The comparison of the dilepton invariant mass $M_{\mu^+ \mu^-}$ and 
$M_{e^+e^-}$ 
distribution in the case of a genuine three-body decay. \label{Muuoff}}
\end{figure}

We compare the dilepton invariant mass $M_{\mu^+\mu^-}$ and  $M_{e^+e^-}$ 
distributions in Fig. \ref{Muuoff}, where the tree- and one-loop-level 
results, the blow-up of the endpoint region and the relative one-loop 
corrections are shown. From these figures one obtains that the shapes of the 
$M_{\mu^+\mu^-}$ and $M_{e^+e^-}$ distributions are identical 
at tree level and different at one-loop level
due to the different treatment of collinear-photon radiations. 
In contrast to the numerical results from the SPS1a parameter set 
(see Fig. \ref{SPS1aMuu}), the mass effect is small in Fig. \ref{Muuoff}, 
but it is still distinct, especially in the relative one-loop corrections.
In the calculations for the invariant mass distribution, the momentum of a 
collinear photon is added to that of the emitting electron,
but it is not added to that of the emitting muon. Hence the invariant mass 
$M_{\mu^+\mu^-}$ is reduced in comparison with $M_{e^+e^-}$. 
It leads to the shifting of events from the upper invariant-mass region to 
the lower invariant-mass region. This effect can be seen in the lower frames 
in Fig. \ref{Muuoff}, i.e. in the lower invariant-mass region the relative 
one-loop corrections of the $\mu^+\mu^-$ final state
is larger than that of $e^+e^-$ final state, while the inverse relation holds  
in the upper invariant-mass region.

The decay width of different $\tilde{\chi}_2^0$ decay modes and the branching
ratios of its visible decays are shown in Table~\ref{threebodytotal}. The
total $\tilde \chi_2^0$ decay width is about 600 times smaller than
for scenario SPS1a. This is not surprising, since
in this scenario $\tilde \chi_2^0$ can only have three-body decays,
while the two-body decays $\tilde \chi_2^0 \rightarrow \tilde l_1^\pm 
l^\mp \to \tilde \chi_1^0 l^-l^+$ are kinematically allowed in the scenario 
SPS1a. In the modified SPS1a scenario one-loop corrections increase the total
$\tilde \chi_2^0$ decay width by a modest 1.2\%.

\begin{table}[hpt]
\begin{center}
\begin{tabular}{|l|l|l|}\hline
decay mode & tree-level width(keV), Br& 1loop-level width(keV), Br\\ \hline
$e^{-} e^{+}\tilde{\chi}_1^0$ &$1.270$, \hspace*{1.8cm}\ \ \ \ \ $4.4\%$ &
$1.451$, \hspace*{1.8cm}$5.0\%$  \\ \hline 
$\mu^{-} \mu^{+}\tilde{\chi}_1^0$ & $1.270$, \hspace*{1.8cm}\ \ \ \ \ $4.4\%$ &
$1.451$, \hspace*{1.8cm}$5.0\%$\\ \hline 
$\tau^{-} \tau^{+}\tilde{\chi}_1^0$ & $7.209$, \hspace*{1.8cm}\ \ \ \ $25.1\%$
&
$7.383$, \hspace*{1.8cm}$25.4\%$\\ \hline 
$\nu_e \bar \nu_e \tilde{\chi}_1^0$ & $1.273$ \hspace*{1.8cm}$$ &
$1.355$ \hspace*{1.8cm}$$\\ \hline 
$\nu_\mu \bar \nu_\mu \tilde{\chi}_1^0$  & $1.273$ \hspace*{1.8cm}$$ &
1.355 \hspace*{1.8cm}$$  \\ \hline 
$\nu_\tau \bar \nu_\tau \tilde{\chi}_1^0$ & $1.273$  \hspace*{1.8cm}$$ &
$1.354$  \hspace*{1.8cm}$$ \\ \hline 
$u \bar u \tilde{\chi}_1^0$ & $2.480$  \hspace*{1.8cm}$$ &
$2.386$  \hspace*{1.8cm}$$ \\ \hline 
$d \bar d \tilde{\chi}_1^0$ & $3.330$  \hspace*{1.8cm}$$ &
$3.298$  \hspace*{1.8cm}$$ \\ \hline 
$c \bar c \tilde{\chi}_1^0$ & $2.475$  \hspace*{1.8cm}$$ &
$2.378$  \hspace*{1.8cm}$$ \\ \hline 
$s \bar s \tilde{\chi}_1^0$ & $3.330$  \hspace*{1.8cm}$$ &
$3.298$  \hspace*{1.8cm}$$ \\ \hline 
$b \bar b \tilde{\chi}_1^0$ &3.595  \hspace*{1.8cm}$$ &
$3.405$  \hspace*{1.8cm}$$ \\ \hline 
total width & $28.778$& $29.114$\\ \hline
\end{tabular}
\end{center}
\caption{The decay width of different $\tilde{\chi}_2^0$ decay modes and the
  branching ratios of its visible leptonic decays in the modified SPS1a
  scenario. \label{threebodytotal}}
\end{table}

Turning to the various partial widths of leptonic decays, we notice that the 
$\tilde \tau^+\tilde \tau^- \tilde \chi_1^0$ final state is still the largest 
decay mode of $\tilde \chi_2^0$ ($25.1\%$ at tree level, $25.4\%$ at one-loop 
level) since $m_{\tilde \tau_1}$ is smaller than the selectron and smuon
masses and the large $L-R$ mixing exists only in the $\tilde \tau$ sector. Note 
that exchange of the $SU(2)$ doublet sleptons now dominates for $l=e, \, \mu$
since the size of the $\tilde \chi_2^0 \tilde e_L e$ coupling exceeds that of
the $\tilde \chi_2^0 \tilde e_R e$ coupling by nearly a factor of 10
and all the sleptons are off shell in this scenaro. 
This dominance of $\tilde e_L$ exchange also explains why the $e^+ e^- \tilde
\chi_1^0$ and $\nu_e \bar \nu_e \tilde \chi_1^0$ final states now have quite
similar partial widths: in the limit where $\tilde \chi_2^0$ and $\tilde
\chi_1^0$ are pure $SU(2)$ and $U(1)_Y$ gauginos, respectively, the product of
couplings involved in $\tilde e_L$ and $\tilde \nu_e$ exchange is exactly the
same (up to an overall sign).

The pattern of one-loop corrections to leptonic decays is different 
from the original SPS1a scenario. The partial width into electrons and muons 
is still enhanced by about 14.3\%. But now the invisible partial widths are 
also increased, diminishing the correction of the branching rations. 
In the original SPS1a scenario, 
the one-loop partial widths of the invisible decays are almost unchanged in
comparison with the tree-level ones, see Table \ref{twobodytotal}.
Note that we assumed three exactly degenerate sneutrinos here, unlike in the
original SPS1a scenario, where $\tilde \nu_\tau$ is slightly lighter than
$\tilde \nu_e$. In the modified scenario a tiny difference between the partial
widths for $\nu_\tau \bar \nu_\tau \tilde \chi_1^0$ and $\nu_e \bar \nu_e
\tilde \chi_1^0$ final states nevertheless results from one-loop corrections
involving the $\tau$ mass or Yukawa coupling (e.g. from the $\tilde \nu$ and
$\nu$ two-point functions).

The hadronic final states have very large partial decay widths and branching 
ratios: $\Gamma_{\rm hadronic}^{\rm tree} = 15.210$ keV ($52.9\%$), 
$\Gamma_{\rm hadronic}^{\rm 1-loop} = 14.765$ keV ($50.7\%$), 
though the squark masses are much larger than the slepton masses.
Part of the reason is that the $Z$-exchange diagrams give larger contributions 
to hadronic final states than to leptonic ones. Moreover, the interference 
between $Z$ and sfermion exchanges is large and positive for the hadronic 
final states, while it is also 
large but negative for the leptonic final states. This is the main reason why 
the hadronic decays of $\tilde \chi_2^0$ obtain so large branching ratios.

\section{Summary and conclusions}\label{conclusion}

We have performed a complete one-loop calculation of the decays
$\tilde{\chi}_2^0 \rightarrow l^{-} l^{+} \tilde{\chi}_1^0 \ (l = e,\, \mu,\,
\tau)$. The necessary renormalization is briefly described in Sec.~2. In most
cases we used on-shell renormalization, which leaves the masses of the
relevant neutralinos and sleptons (almost) unchanged. This is convenient for
our purpose, since one important goal in the experimental study of leptonic
$\tilde \chi_2^0$ decays is the determination of (differences of)
supersymmetric particles masses from the dilepton invariant mass 
($M_{l^+ l^-}$)
distribution.

For the cases where the intermediate charged sleptons can be on 
shell, these decays were calculated both completely and in a single-pole 
approximation at one-loop level. In the complete calculation one has to 
employ complex slepton masses in the relevant propagators and one-loop 
integrals. The single-pole approximation in this case is performed in 
the way that the $\tilde{\chi}_2^0$ decays are treated as a sequence of 
two two-body decays. We checked that for the well-studied SPS1a
parameter set, this approximation reproduces the integrated partial widths to
better than 0.5\% accuracy even after one-loop corrections are included. 
For this parameter set we find a rather small one-loop correction
to the total $\tilde \chi_2^0$ decay width, but the branching ratio for the
most easily detectable electron and muon final states are increased by about
13.6\% at one-loop level.

We also studied the effect of higher-order corrections on the $M_{l^+l^-}$
distribution. If only one exchanged particle can be on-shell, as in the SPS1a
scenario, the shape of this distribution is altered only by real photon
emission contributions, i.e. its peak is shifted by several GeV below the 
endpoint. This is very important since the shape of the distribution near 
the endpoint should be known if the endpoint is to be determined accurately 
from real data. In our calculation we define collinear photons as being
emitted at an angle $\Delta \theta < 1^\circ$ relative to the emitting lepton.
Since the selectrons and smuons have equal masses and the light lepton mass 
$m_l~(l = e, \mu)$ is neglected except when it appears in the one-loop 
integrals, one will obtain identical distributions for $M_{e^+e^-}$ and 
$M_{\mu^+ \mu^-}$ if the momentum of a collinear photon is added to that of 
the emitting lepton. The actual effect of the collinear-photon radiation 
depends on details of the measurement apparatus, and therefore has to be 
calculated anew for each experiment.
We have focused on the LHC experiment in our calculation.
At the LHC the electron energy is determined calorimetrically.
In this case a collinear photon would hit the same cell of the calorimeter as
the electron, so the two energies cannot be disentangled. Hence
we add the momentum of a collinear photon to the one of the emitting electron 
in our calculation. 
Since muons pass through the calorimeter, where the photons are detected, and 
measured further outside in the muon detector at the LHC 
(their 3-momenta are measured through the curvature radius of their track in 
the magnetic field), the momentum of a collinear photon is not added to the 
one of its emitter muon in our calculation. In this case the mass effect can 
be seen in the dilepton invariant mass distribution. We find that the peak of 
the $M_{e^+e^-}$ distribution is moved downwards by about $4$ GeV once the 
one-loop corrections are added. In comparison to the $M_{e^+e^-}$ distribution, 
the peak of the $M_{\mu^+\mu^-}$ distribution is shifted slightly to lower 
invariant-mass values at one-loop level. 
This is due to the different treatment of the collinear-photon radiation.

We have also analyzed a modified SPS1a scenario, with increased slepton
masses, so that $\tilde{\chi}_2^0$ can only undergo genuine three-body
decays. In this case we again find a moderate, if slightly larger, correction
to the total $\tilde \chi_2^0$ width when one-loop corrections are considered, 
but the branching ratios for the electron and muon final states are still
enhanced by about 13.6\% at the one-loop level. We have seen in 
Fig.~\ref{universal} that for the simpler case $l=e, \mu$ the bulk of 
the non-QED correction to the partial width can be absorbed into new 
$\tilde \chi_2^0 \tilde l_1 l$ couplings, which are sensitive to the 
spectrum of sfermions. In the case of $\tau$
final states, significant $\tilde \tau_L - \tilde \tau_R$ mixing as well as
the $\tau$ Yukawa coupling have to be included in the analysis. We have not
attempted to define such effective couplings and, perhaps, mixing angles here.

In this modified SPS1a scenario, the dilepton invariant mass distributions
have a rather complicated shape, showing the contributions from $Z$
exchange near the upper endpoints of the distributions. In this case the shape
of these distributions is affected not only by real photon emission, which
again leads to significant negative corrections for large $M_{l^+l^-}$, but
also by virtual corrections, which can e.g. differ for $Z$ and slepton
exchange diagrams. In this case the shape of the distribution away from the
endpoint also carries information about slepton masses and neutralino mixing
angles. Fitting tree-level distributions to real data might therefore give
wrong results for these physical parameters. In this context a careful
analysis of collinear radiation is also important, since differences in the
energy measurements of electrons and muons could lead to spurious differences
of fitted selectron and smuon masses. Here the collinear-photon radiations 
for electrons and muons are treated as discussed beforehand.
One finds that the one-loop shapes of the $M_{e^+e^-}$ and $M_{\mu^+\mu^-}$ 
distributions are different, though the selectrons and smuons have equal 
masses in our calculations.

We conclude that higher-order corrections to leptonic $\tilde \chi_2^0$ decays
can exceed the 10\% level both in integrated partial widths and branching
ratios, and in the shape of the dilepton invariant mass distribution.
Attempts to absorb much of the large virtual corrections into effective
running couplings might be rewarding. An accurate understanding of $\tilde
\chi_2^0$ decays is of considerable importance, since this is one of the
lightest visible particles that can be produced directly at future $e^+e^-$
colliders, and plays a prominent role in the analysis of cascade decays of
gluinos and squarks at the LHC.

\section*{Acknowledgements}
We would like to thank T. Fritzsche, T. Hahn, and H. Rzehak for 
helpful discussions, as well as A. Bredenstein and M. Roth for help with 
calculating the hard photon bremsstrahlung contribution. 
MD thanks the School of Physics at KIAS, Seoul, as well as the
theory group at the university of Hawaii for hospitality.
\newpage
\section*{Appendix}

\begin{appendix}

\section{Parameters}\label{Parameter}

For the numerical evaluation, the following values of the SM parameters are
used:
\begin{eqnarray}
m_e &=& 0.510999 {\rm MeV}\, , \hspace*{3mm}m_\mu = 105.6584 {\rm MeV}\,  , 
\hspace*{3mm}
m_\tau =   1.777{\rm GeV}\ , \nonumber \\ 
m_u &=& 53.8 {\rm MeV}\, ,  \hspace*{1.1cm}m_c =  1.5 {\rm GeV}\, , 
\hspace*{14mm}m_t =   175{\rm GeV} \, ,\nonumber \\ 
m_d &= & 53.8 {\rm MeV}\, ,  \hspace*{1.1cm}m_s = 150 {\rm MeV}\, , 
\hspace*{12mm}m_b = 4.7 {\rm GeV}\, , \nonumber \\
 m_W & = & 80.45{\rm GeV} ,  \hspace*{10mm} m_Z = 91.1875{\rm GeV}\, ,
\nonumber 
\\
\alpha(0) & = &1/137.0359895, \hspace*{10mm}G_\mu = 1.1663910\times 10^{-5}
{\rm GeV}^{-2} \, . \nonumber 
\end{eqnarray}
The on-shell renormalization scheme requires
$\alpha =\alpha(0)$ for one-loop calculations.
For the tree level expressions we instead 
use the effective coupling for the overall normalization,
\begin{eqnarray}
\alpha_{G_\mu} & = & \frac{\sqrt{2}G_\mu M_W^2 s_W^2}{\pi}\, .
\end{eqnarray}
We saw in Sec.~5 that this choice leads to good perturbative stability of the 
total $\tilde \chi_2^0$ decay width.

\end{appendix}

\end{document}